%% file: main.tex
\documentclass[aps,pre,twocolumn,groupedaddress,floatfix,longbibliography]{revtex4-2}

\usepackage{amsmath,amssymb,amsthm,braket,listings,physics,
            here,subfigure,longtable,bbm,mathtools,xcolor,bm}
\usepackage[normalem]{ulem}
\usepackage{graphicx}
\usepackage[hidelinks]{hyperref}
\usepackage[sectionbib]{bibunits}
\usepackage{autobreak}

\newtheorem{proposition}{Proposition}
\newtheorem{corollary}{Corollary}

\newtheorem{assumption}{Assumption}

\newcommand{\pt}{\partial}
\newcommand{\relmiddle}[1]{\mathrel{}\middle#1\mathrel{}}
\newcommand{\rotatedsquare}{\mathbin{\text{\rotatebox[origin=c]{45}{$\blacksquare$}}}}

\DeclareMathOperator{\argmax}{argmax}
\newif\ifshowcomments
\showcommentsfalse 

\makeatletter
\newcommand{\HideDraftAnnotation}{\@bsphack\@esphack}
\makeatother

\definecolor{DraftNoteColor}{HTML}{A66B00}
\definecolor{EnglishEditColor}{HTML}{228B22}
\ifshowcomments
\newcommand{\DraftNote}[1]{\textcolor{DraftNoteColor}{#1}}
\newcommand{\Kato}[1]{\textcolor{red}{Comment: #1}}
\newcommand{\Reply}[1]{\textcolor{blue}{Reply: #1}}
\newcommand{\TODO}[1]{\textcolor{teal}{TODO: #1}}
\newcommand{\Kobayashi}[1]{\textcolor{blue}{Kobayashi: #1}}

\newcommand{\DEL}[1]{{\color{red}\sout{#1}}}
\else
\newcommand{\DraftNote}[1]{\HideDraftAnnotation}
\newcommand{\Kato}[1]{\HideDraftAnnotation}
\newcommand{\Reply}[1]{\HideDraftAnnotation}
\newcommand{\TODO}[1]{\HideDraftAnnotation}
\newcommand{\Kobayashi}[1]{\HideDraftAnnotation}

\newcommand{\DEL}[1]{\HideDraftAnnotation}
\fi

\usepackage{ulem}
\usepackage{xcolor}
\usepackage {cancel}
\usepackage{CJKutf8}
\usepackage{ifthen}
\ifshowcomments
\newcommand{\COMM}[2]{{
\begin{CJK}{UTF8}{ipxm}
\ifthenelse{\equal{#1}{TJK}}{\color{red}}{
\ifthenelse{\equal{#1}{SN}}{\color{blue}}{
\ifthenelse{\equal{#1}{MK}}{\color{cyan}}{
\ifthenelse{\equal{#1}{BB}}{\color{magenta}}}}}
[#1: #2]
\end{CJK}
}}
\else
\newcommand{\COMM}[2]{\HideDraftAnnotation}
\fi

\makeatletter
\renewcommand\appendix{\par
   \setcounter{section}{0}%
   \renewcommand{\thesection}{\Alph{section}}
   \renewcommand{\thesubsection}{\alph{subsection}}
}
\makeatother

\begin{document}


\title{Chemotactic aggregation as cue-mediated adaptation \protect\\ toward Nash equilibrium in a spatial patterning game}


\author{Masaki Kato${}^{1}$}
\email{m\_kato@sat.t.u-tokyo.ac.jp}
\author{Tetsuya J. Kobayashi${}^{2,1,3}$}
\email{tetsuya@mail.crmind.net}
\affiliation{${}^{1}$Department of Mathematical Informatics, the Graduate school of Information Science and Technology, The University of Tokyo, Tokyo, Japan}
\affiliation{${}^{2}$Institute of Industrial Science, The University of Tokyo, Tokyo 153-8505, Japan}
\affiliation{${}^{3}$Universal Biology Institute, The University of Tokyo, Tokyo 113-8654, Japan}


\date{\today}

\begin{abstract}
\DraftNote{You can show the editing notes and comments by setting \texttt{\textbackslash showcommentstrue} and hide them by setting \texttt{\textbackslash showcommentsfalse} around l.72 in \texttt{main.tex}.}
\DraftNote{Notation conventions are specified in the preamble of main.tex.}
\DraftNote{[Cn] represents the logical elements consitituting this draft paper. Details in \texttt{CLAIM\_MAP.md}}

\DraftNote{[C1]}
Chemotactic cells generate their own chemical gradients to form spatial patterns that support collective functions ranging from cooperative metabolism to immunity. 
\DraftNote{[C2]}
Despite progress in understanding the optimality of chemotaxis in imposed chemical gradients, 
it remains unclear how close decentralized cell populations can come to maximizing the population-wide benefit of spatial organization through individual movement and local modulation of chemical gradients.
\DraftNote{[C5]}
We employ mean field game theory to characterize efficient spatial patterns achievable by decentralized populations as Nash equilibria:
each agent optimizes its positioning given the spatial density of the other cells, whereas the population density is determined by the positioning of all individuals.
\DraftNote{[C8][C10]}
We derive coupled dynamics of population movement and cue modulation that drive adaptation toward these equilibria and verify its structural correspondence to established phenomenological models of chemotactic pattern formation.
\DraftNote{[C11][C12][C13]}
We apply our framework to self-organized aggregation and numerically demonstrate that the dynamics can form multiple equilibrium patterns, some of which are comparable to the population-level optimum under moderate incentives to aggregate.
\DraftNote{[C14]}
We also show that the relative timescales of cue modulation and movement influence pattern selection and population-level efficiency:
slower cue modulation can yield more coherent and efficient equilibrium patterns.
\DraftNote{[C15]}
Our framework thus highlights fundamental limits on collective optimization in decentralized cell populations, and suggests a functional role for timescales in self-organized coordination through environmental feedback.
\end{abstract}


\maketitle


\renewcommand{\theequation}{\arabic{equation}}

\input{1-introduction.tex}
\input{2-model.tex}
\input{3-experiment.tex}

\input{4-conclusion-discussion.tex}

\begin{acknowledgments}
The authors thank Simon K. Schnyder, Takehiro Tottori and Shuhei A. Horiguchi for valuable discussions.
The first author received the JSPS Research Fellowship Grant No.24KJ0846 and was partially supported by the World-leading Innovative Graduate Study Program in Proactive Environmental Studies (WINGS-PES), The University of Tokyo. 
This work was supported by JSPS Grant-in-Aid for Transformative Research Areas (A) (25H01365) and JST CREST (JPMJCR25Q2). 
\end{acknowledgments}

\bibliography{reference.bib}%


\onecolumngrid
\appendix 
\clearpage
\setcounter{section}{0}

\renewcommand\appendixname{Supplement}

\section{Theoretical foundations and analysis}
\setcounter{page}{1}
\setcounter{equation}{0}
\setcounter{figure}{0}
\renewcommand{\theequation}{S\arabic{equation}}
\renewcommand{\thefigure}{S\arabic{figure}}
\renewcommand{\thepage}{S\arabic{page}}

\input{AppendixA-Theory}
\input{AppendixB-Deriv-Limit}
\input{AppendixC-SupplementForExperiments}



\end{document}

%% file: 1-introduction.tex
\section{Introduction}
\begin{figure*}[tbp]
    \hfill
    \begin{minipage}[b]{\linewidth}
        \centering
    \includegraphics[width=\textwidth]{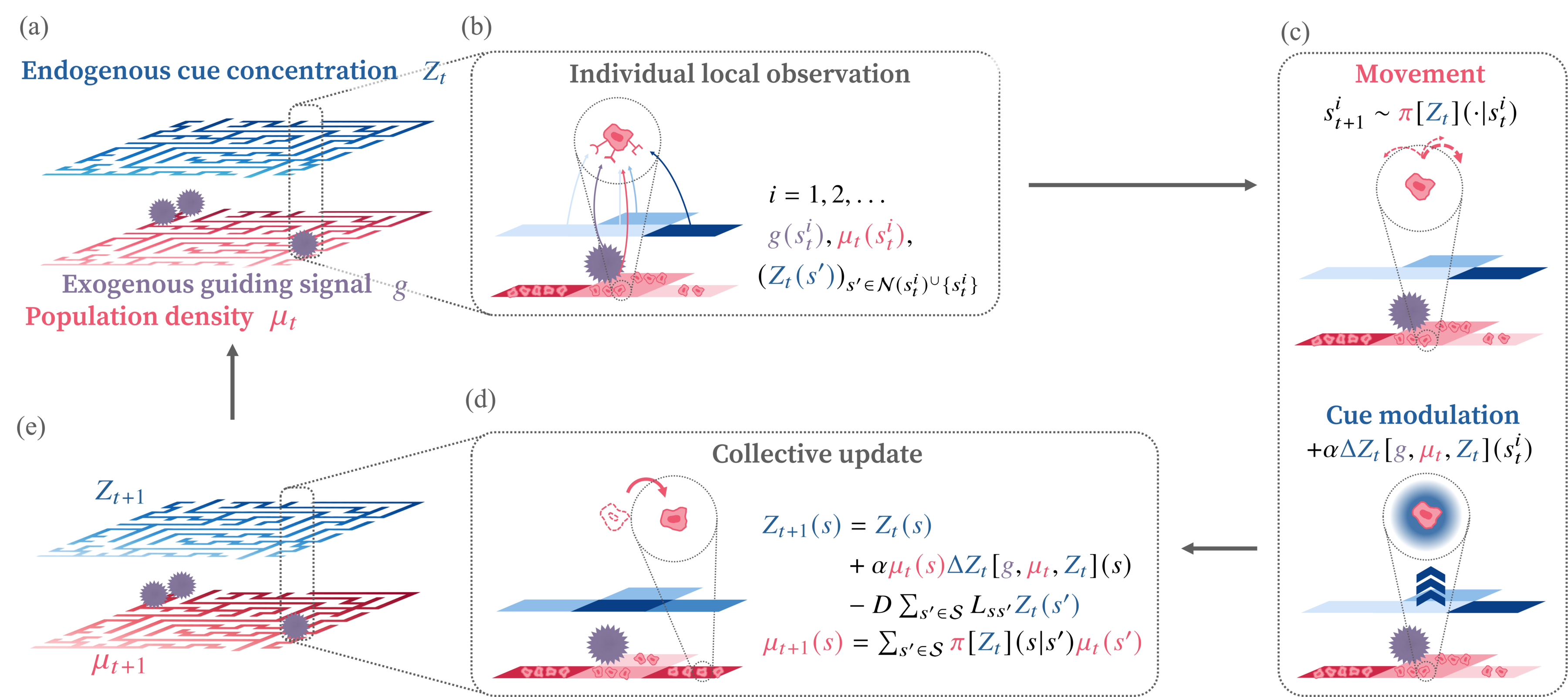}
    \end{minipage}
    \hfill
    \caption{
        Schematic of an agent population that forms spatial patterns through endogenous cue modulation.
        (a) A large number of agents move in an environment, with their spatial distribution represented by $\mu_t$. 
        Three quantities are defined at each vertex $s$ of the environment: population density $\mu_t(s)$, a fixed exogenous guiding signal $g(s)$, and an endogenous cue $Z_t(s)$ altered by agents.
        (b) At each time $t$, each agent $i=1,2,\ldots$ located at vertex $s_t^i$ observes the local exogenous signal $g(s_t^i)$, the population density $\mu_t(s_t^i)$, and nearby endogenous cue concentrations $(Z_t(s'))_{s'\in\mathcal{N}(s_t^i)\cup\{s_t^i\}}$.
        (c) Based on these local observations, each agent moves between vertices according to movement policy $\pi[Z_t]$ and modulates the local cue concentration by $\Delta Z_t[g,\mu_t,Z_t](s_t^i)$ at rate $\alpha$.
        $\Delta Z_t<0$ corresponds to degradation, and $\Delta Z_t>0$ corresponds to production.
        The movement policy $\pi[Z_t]$ and cue modulation rule $\Delta Z_t[g,\mu_t,Z_t]$ defined in Eqs.~\eqref{eq:dist-G-learning-policy} and \eqref{eq:dist-G-learnSMFG-diff} allow agents to adapt toward individually optimal movement, enabling the population to approach equilibrium patterns.
        (d) Individual movement redistributes the population, while the combined cue modulation of local agents changes the cue concentration.
        The endogenous cue also diffuses independently of the agents with diffusion coefficient $D$.
        (e) The updated population density $\mu_{t+1}$ and cue field $Z_{t+1}$ determine the agents' subsequent local observations.
    }
    \label{fig:schematics_model}
\end{figure*}

\DraftNote{[Functional pattern]}
\DraftNote{[C1]}
Spatial patterns are ubiquitous in cellular systems --- ranging from bacteria~\cite{Budrene1991Nature,Davies1998Science} through social amoebae~\cite{Bonner2009} to human cells~\cite{Aceto2014Cell,Lammermann2013Nat}--- and 
enable collective functions beyond individual capabilities such as cooperative metabolism~\cite{Shapiro1998AnnuRevMicrobiol,Evans2020TrendsMicrobiol}, collective protection~\cite{Stewart2008NatRevMicrobiol,Oliveira2016NatRevImm,Flemming2016NatRevMicroBiol,Jo2022NatRevMicrobiol} and long-range dispersal~\cite{Bonner2009,Claessen2014NatRevMicrobiol}.
\DraftNote{[Patterns through chemotaxis and self-generated gradient]}
Such patterning is often mediated by intercellular communication via extracellular chemical cues, both in the presence~\cite{Lammermann2013Nat, Long2017SciRep,Tweedy2020Science} and absence~\cite{Budrene1991Nature,Budrene1995Nature,Davies1998Science,Ben-Jcob2000,Mittal2003PNAS,Park2003PNAS,Laganenka2016NatComm} of exogenous guiding signals, i.e. signals from the environment.
For instance, bacterial populations can self-organize into aggregates and biofilms by modulating endogenous, i.e. their own, chemical cues depending on local cell density~\cite{Budrene1991Nature,Budrene1995Nature,Davies1998Science,Ben-Jcob2000,Mittal2003PNAS,Park2003PNAS,Laganenka2016NatComm}.
Immune cells can effectively accumulate around injury sites by producing endogenous chemical cues that amplify faint exogenous guiding signals from injuries
while degrading the endogenous cues in distant regions~\cite{Insall2022TrdCellBiol}.
Greater accumulation is not always beneficial, as excessive accumulation can cause inflammation~\cite{Glaser2021CurrOpiCellBiol} or intensify competition for resources~\cite{Stewart2008NatRevMicrobiol,Claessen2014NatRevMicrobiol}.
Cell populations regulate their density and limit excessive accumulation, helping maintain functional spatial patterns.

\DraftNote{[Plan II: Functional patterns through local responses]}
\DraftNote{[C1][C3]}
These functional spatial patterns emerge as individual cells adjust their movement and modulate endogenous chemical cues in response to locally sensed cell density, guiding signals and cues.
A fundamental question for cell populations whose spatial patterns emerge from local responses is how close these populations can come to the population-level optimum of spatial organization~\cite{Guo2018PNAS,Bruckner2026PRXlife,Tripathi2026PRXLife,Arzash2026PNAS}.
For example, a study on bacterial colonies investigates whether local regulation of enzyme expression can produce spatial expression patterns with optimal metabolic efficiency~\cite{Guo2018PNAS}.
In motile cell populations, however, the efficiency of spatial patterns achievable through individual movement and cue modulation remains poorly understood.

\DraftNote{[C2]}
The efficiency of biological behaviors can be assessed using normative models, in which behavior is derived by optimizing a performance metric representing the biological system's goals.
Normative models have been used to understand efficient cellular information processing including single-cell chemotaxis~\cite{Clark2005PNAS,Celani2010PNAS,Mattingly2021NatPhys,Nakamura2022PRR,Mattingly2026NatPhys}.
Normative studies of cell populations that communicate through extracellular chemicals have likewise focused primarily on the limits of sensing external chemical gradients~\cite{Ellison2016PNAS,Fancher2017PRL,Saha2026PRE}.
Spatial pattern formation, however, has been studied mainly through mechanistic models~\cite{Keller1970JTheorBiol,Tsimring1995PRL,Woodward1995BiophysJ,Brenner1998BiophysJ,Tyson1999ProcRSocLondB,Painter2019JTB}, which directly describe underlying processes, while normative frameworks for evaluating function and efficiency of these patterns remain limited. 
Our previous work~\cite{Kato2025PRXLife} addressed this gap for the specific case of accumulation near sources of exogenous guiding signals.
We identified how cue modulation should be coupled to chemotactic movement to achieve efficient accumulation.
However, that framework did not account for situations in which the desirability of a location for a cell depends on the distribution of other cells, as in self-organized aggregation.

\DraftNote{[C3]}
This interdependence raises additional challenges for normative modeling of spatial pattern formation.
\DraftNote{[C3]}
First, cells adjust their movement without a central controller, so the resulting spatial patterns need not maximize population-level efficiency even when each cell optimizes its own movement.
A pioneering study derived an efficient dispersal strategy for a decentralized homogeneous population~\cite{Pezzotta2018PRE}.
The strategy can be interpreted as a Nash equilibrium: each cell moves to maximize its own outcome given the population density generated by the other cells following the same strategy.
However, that study addressed neither aggregation nor the dynamics leading to equilibrium.
In aggregation, local aggregates can be sustained by cells that cannot improve their own outcomes by unilaterally changing their movement.
Such aggregates may differ in their density profiles and occur in different spatial arrangements, allowing alternative equilibrium patterns with different population-level efficiencies.


\DraftNote{[C14]}
Second, a further issue is how the population comes to form a particular equilibrium pattern.
Spatial patterns form through feedback between population movement and the environmental cue: individual cells modulate local cue concentration depending on the local population density, while their cue-guided movement in turn changes that density (Fig.~\ref{fig:schematics_model}). 
With such feedback, the relative timescales of movement and cue modulation determine how much the population redistributes while the cue field adjusts, potentially changing the subsequent course of aggregation.
In decentralized populations with environmental feedback, relative timescales have been shown to affect which equilibrium is reached and hence the collective outcome~\cite{Shao2019EPL,Hauert2019JTheorBiol,Tilman2020NatCommun}.
These physical and theoretical considerations raise the question of how the relative timescales of movement and cue modulation influence the morphology and population-level efficiency of the resulting spatial patterns.


\DraftNote{[C5]}
To address these issues, we characterize efficient spatial patterns achievable by decentralized cell populations as Nash equilibria of a stationary mean field game (SMFG).
At equilibrium, each cell optimizes its movement policy given the population density, while the population's movement leaves that density unchanged.
\DraftNote{[C8]}
We derive dynamics of chemotactic agents that approach these equilibrium patterns through endogenous cue modulation:
agents cooperatively modulate the cue based on local population density to improve their own cue-guided movement policies (Fig.~\ref{fig:schematics_model}).
\DraftNote{[C8][C10]}
These dynamics structurally correspond to the phenomenological Keller-Segel type models, suggesting that chemotactic populations may approach these equilibrium patterns.
\DraftNote{[C11][C12][C13]}
We apply this framework to self-organized aggregation of chemotactic populations and 
numerically demonstrate that the dynamics can form multiple equilibrium patterns depending on initial densities.
Some of these patterns are comparable to the population-level optimum under moderate incentives to aggregate.
\DraftNote{[C14]}
We also find that the relative timescales of movement and cue modulation affect pattern selection and population-level efficiency. 
Slower cue modulation can lead to more coherent patterns with higher population-level efficiency, possibly by allowing redistribution before cue-mediated feedback reinforces local accumulation.
Our framework thus highlights constraints on collective optimization in decentralized cell populations and the role of timescales in shaping their collective performance.

%% file: 2-model.tex
\section{Model}
\label{sec:model-intro}

\subsection{Formulation of spatial patterning by chemotactic agents}
\label{sec:RL-formulation}
\DraftNote{[C4]}
To model spatial patterning by chemotactic cells, we consider a population of agents moving in an environment. 
We adopt the discrete-time discrete-space formulation of our previous work~\cite{Kato2025PRXLife} for simplicity.
We represent the space in which agents move as a finite connected undirected graph $\mathcal{G} = (\mathcal{S}, \mathcal{E})$ with vertex set $\mathcal{S}$ and edge set $\mathcal{E}$, including lattices as a special case.
We describe a sufficiently large population of indistinguishable agents by a probability density $\mu_t\in\Delta(\mathcal{S})$ at time step $t\in\{0,1,\ldots\}$. 
The density $\mu_t(s)$ represents the proportion of agents at vertex $s$, with multiple agents allowed to occupy the same vertex.
For a representative agent $i\in\mathbb{N}$ in the population, we denote its location at time $t$ by $s^i_t\in\mathcal{S}$.
At each time step $t$, agent $i$ stochastically chooses either to stay at $s'=s_t^i$ or to move to a neighboring vertex $s'\in \mathcal{N}(s_t^i)$ with probability $\pi_t^i(s'|s_t^i)$, 
where $\mathcal{N}(s)$ is the set of neighboring vertices of $s$. 
We call the transition rule $\pi_t^i:\mathcal{S}\times\mathcal{S}\to[0,1]$ the agent's movement policy.

\DraftNote{[C4]}
When forming aggregates, cells adapt their motility depending on their local environment.
An important element is the number of neighboring cells.
Cells can adjust their responses depending on the number of neighboring cells~\cite{Kolenbrander2002MicrobiolMolBiolRev,Laganenka2016NatComm,Strickland2024DevCell}.
By sensing neighbors indirectly through chemical signals, cells may estimate how many are present without distinguishing individual neighbors~\cite{Yuen1994JTheorBiol}.
Thus, we assume that an agent at vertex $s$ can sense the local density $\mu_t(s)$ with sufficient precision for simplicity. 

\DraftNote{[C4]}
Another critical factor for aggregation is environmental heterogeneity such as resource availability and risk of harm.
\textit{E. coli} and neutrophils sense exogenous guiding signals such as glucose~\cite{Ha2018SciAdv,Scheidweiler2024NatCommun} and fMLP from damaged tissue~\cite{Szatmary2017MolBiolCell}, respectively.
We represent such environmental heterogeneity of exogenous guiding signals by a fixed spatial distribution $g\in\mathbb{R}^{\mathcal{S}}_{\ge0}$.
While exogenous guiding signals may be consumed by agents~\cite{Brenner1998BiophysJ}, we assume for simplicity that they are not altered by the agents.



\DraftNote{[C4]}
Primary guiding signals from neighboring cells and the environment are often too weak for cells to aggregate effectively.
Cells coordinate their movement by secreting diffusible endogenous chemical cues that amplify local signals and relay information to distant cells~\cite{Kolaczkowska2013NatRevImm,Ng2011JID,Afonso2012DevCell,Glaser2021CurrOpiCellBiol,Laganenka2016NatComm}.
For example, \textit{E. coli} and \textit{Dictyostelium} produce or take up attractant depending on environmental conditions or local cell density, contributing to biofilm formation or aggregation~\cite{Laganenka2016NatComm,Ha2018SciAdv,Scheidweiler2024NatCommun,Yuen1995JCellBiol}.
Neutrophils likewise secrete endogenous cues such as LTB4 to promote recruitment to damaged tissue~\cite{Ng2011JID,Afonso2012DevCell,Glaser2021CurrOpiCellBiol}.
Cue modulation may depend not only on the exogenous guiding signal but also on the local cell density~\cite{Poplimont2020CurrBiol}, potentially amplifying recruitment.
Neutrophils also have mechanisms to limit excess accumulation, which can cause inflammation and tissue damage~\cite{Hopke2020NatCom,Kienle2021Science,Mihlan2022FrontCellDevBiol}.
Density-dependent suppression of cue secretion could be one such mechanism.
More generally, density-dependent cue modulation could allow aggregating cell populations to regulate their local density.
For simplicity, we consider a single endogenous cue, whose spatial distribution is represented by $Z_t:\mathcal{S}\to\mathbb{R}_{\geq 0}$ (Fig.~\ref{fig:schematics_model} (a)).

\DraftNote{[C4][C7]}
Finally, we assume that agents guide their movement using only current local observations, without internal memory.
Specifically, each agent's movement policy $\pi_t^i$ depends only on the exogenous guiding signal $g(s^i_t)$, the population density $\mu_t(s^i_t)$ at its current vertex $s_{t}^{i}$, and the endogenous cue at that vertex and its neighbors $(Z_t(s))_{s\in \{s^i_t\}\cup \mathcal{N}(s^i_t)}$ (Fig.~\ref{fig:schematics_model} (b)).
The endogenous cue, which agents sense and modify, may nevertheless serve as a shared external memory for coordination.

\subsection{Characterization of equilibrium patterns}
\DraftNote{[C3][C5]}
Collective modulation of the endogenous cue need not ensure fully coordinated movement, as each agent still rely on local observations to adjust its movement.
\DraftNote{[C5]}
We therefore model each agent as seeking a better movement policy based on local observations, while treating other agents and the exogenous guiding signal as its environment.
We describe how desirable each vertex $s\in\mathcal{S}$ is to an agent using a reward function $r_g(s;\mu)\in\mathbb{R}$, which depends on population density $\mu\in\Delta(\mathcal{S})$ and the exogenous guiding signal $g\in\mathbb{R}^{\mathcal{S}}$.
To reflect agents’ local observations, we here restrict the reward to depend on local quantities: $r_g(s;\mu)=r_{g(s)}(s;\mu(s))$. 
When site desirability $r_g(s;\mu)$ is independent of $\mu$ and depends solely on the exogenous signal, our formulation recovers the problem of locating sources of exogenous guiding signals studied in our previous work~\cite{Kato2025PRXLife}.  
To represent a population with shared preferences, we assume the same reward function $r_g$ for all agents.

\DraftNote{[C5]}
As mean-field $\mu_t$ can evolve over time as agents move, optimizing with a time-varying mean-field $\mu_t$ yields a nonstationary optimal movement policy, which is difficult for simple organisms to memorize and learn.
Thus agents are assumed to optimize their stationary movement policy against the stationary mean-field $\mu$.
\DraftNote{[C4][C5]}
For a given population density $\mu$, each agent seeks to find the movement policy $\pi^*_\mu\coloneqq\argmax_\pi \mathcal{J}[\pi;\mu]$ that maximizes the following expected discounted cumulative reward $\mathcal{J}[\pi;\mu]$, including a movement control penalty.
\begin{align}
    \label{eq:cost-EMDPs}
    \mathcal{J}[\pi;\mu]&\coloneqq \mathbb{E}\left[\sum_{t=0}^\infty\gamma^tR^\pi_g(s_t,s_{t+1};\mu)\right],\\
    R^\pi_g(s_t,s_{t+1};\mu)&=r_g(s_t;\mu)-\frac{1}{\beta}\log\frac{\pi(s_{t+1}|s_{t})}{p(s_{t+1}|s_t)},
    \label{eq:temp-rewpen}
\end{align}
where the expectation is taken over trajectories generated by movement policy $\pi$, with initial states drawn from $\mu$.
The discount factor $\gamma \in [0,1)$ is an intrinsic parameter common to all agents, and gives greater weight to immediate reward than to future reward.
It would be natural for simple organisms with limited lifespans to prioritize the present over the distant future.
The additional term $\frac{1}{\beta}\log\frac{\pi}{p}$ penalizes deviation of the cue-dependent movement policy $\pi$ from the stationary intrinsic movement policy $p:\mathcal{S}\times \mathcal{S}\to [0,1]$.
While our framework allows $p$ to depend on density $\mu$ in general, as discussed in the Supplemental Material (SM) Sec.~\ref{sec:entropy-SMFG}, we focus on a density-independent $p$ in the main text. 
Specifically, we use a lazy random walk for $p$
to represent the spontaneous random movement of cells in the absence of cues. 
This form of control penalty has been used in bacterial chemotaxis~\cite{Nakamura2022PRR}, immunological learning~\cite{KatoPRR2021}, neutrophil accumulation at injury sites~\cite{Kato2025PRXLife}. 
The inverse control weight $\beta\in(0,\infty)$ controls the penalty for deviating from intrinsic movement.

\DraftNote{[C5]}
Based on the individual optimization against the rest of the population, an equilibrium pattern is defined by a pair $(\hat{\pi},\hat{\mu})$ of a stationary movement policy and a population density satisfying the following conditions:
\begin{subequations}
    \label{eq:SMFG-NE} 
    \begin{align}
        \hat{\pi}=\pi^{*}_{\hat{\mu}}\coloneqq\argmax_\pi \mathcal{J}[\pi;\hat{\mu}], 
        \label{eq:SMFG-NE-optimality}\\
        \hat{\mu}(s)=\sum_{s'\in\mathcal{S}}\hat{\mu}(s')\hat{\pi}(s|s'),\quad s\in\mathcal{S}.
        \label{eq:SMFG-NE-stationarity}
    \end{align}
\end{subequations}
At equilibrium, no agent can improve its own cumulative reward by unilaterally deviating from the movement policy $\hat{\pi}$ (Condition~\eqref{eq:SMFG-NE-optimality}), and the population density $\hat{\mu}$ remains unchanged under $\hat{\pi}$ (Condition~\eqref{eq:SMFG-NE-stationarity}).
This formulation is a variant of an entropy-regularized stationary mean field game (SMFG)
and the equilibrium pattern~\eqref{eq:SMFG-NE} is a Nash equilibrium of the SMFG~\cite{Anahtarci2023DynGamesAppl,Guan2022ACC} (SM Sec.~\ref{sec:entropy-SMFG}).
For local rewards, Nash equilibria can also be characterized through a population-level variational problem, supporting their interpretation as efficient patterns achievable by decentralized agents~\cite{Lasry-Lions2007JJM,Pezzotta2018PRE}.
The movement policy $\hat{\pi}$ at Nash equilibrium in general differs from a population-level optimum called a social optimum $\pi^*\in\operatorname{argmax}_\pi\mathcal{J}[\pi;\mu^\pi]$, where $\mu^\pi$ is the unique stationary distribution of $\pi$.
See SM Sec.~\ref{sec:entropy-SMFG} for details.

\DraftNote{[C6]}
For a fixed population density $\mu$, we can explicitly characterize the optimal movement policy $\pi^*_\mu$.
We define the state value function $V^\pi_\mu(\cdot):\mathcal{S}\to\mathbb{R}$ as the expected cumulative reward of an agent starting at vertex $s$ and following movement policy $\pi$:
\begin{align}
    \label{eq:def-value}
    V^\pi_\mu(s)&\coloneqq \mathbb{E}\left[\sum_{t=0}^\infty\gamma^t R_g^\pi(s_t,s_{t+1};\mu)\relmiddle|s_0=s\right],\quad s\in\mathcal{S}.
\end{align}
We then define the optimal state value $V^*_\mu:\mathcal{S}\to\mathbb{R}$ as $V^*_\mu(s)\coloneqq \max_\pi V^\pi_\mu(s)$, representing the maximum long-term return attainable from vertex $s$. 
The optimal movement policy $\pi_\mu^*$ is then given by:
\begin{equation}
    \pi_\mu^*(s'|s)=\frac{p(s'|s)\exp(\beta \gamma V_\mu^*(s'))}{\sum_{s'\in\mathcal{S}}p(s'|s)\exp(\beta \gamma V^*_\mu(s'))},\quad s,s'\in\mathcal{S}.
    \label{eq:EMDP-greedy-policy}
\end{equation}
\DraftNote{[C6]}
This policy biases the agent's intrinsic movement $p(s'|s)$ toward neighboring vertices with higher $V^*_\mu(s')$. 
This can be interpreted as a biased random walk along the local gradient of $V^*_\mu$. 

\DraftNote{[C7]}
The optimal value $V^*_\mu$ depends on the exogenous guiding signal and the population density at distant locations;
thus an agent cannot determine the optimal value from local observation of these quantities alone. 
Still, agents may approximate the optimal value by storing and updating information from local observations in a shared external memory. 
Following our previous work~\cite{Kato2025PRXLife}, we posit that the endogenous cue $Z_t$ serves as a shared memory encoding an estimate $V_t$ of the optimal value $V_\mu^*$ through $V_t(s)=\mathcal{F}[Z_t](s)$.
For simplicity, we choose a linear mapping $\mathcal{F}[Z_t](s)\coloneqq Z_t(s)/\beta$; logarithmic and other mappings are also possible here, as in our previous work. 

\subsection{Coupled dynamics of population movement and cue modulation toward equilibrium patterns}
\label{sec:learning-dynamics}
\DraftNote{[C8]}
To describe adaptation toward the equilibrium pattern above, we consider coupled dynamics of population movement and cue modulation (Fig.~\ref{fig:schematics_model} (b-d)).
Chemotactic movement is often modeled as a biased random walk along a chemical gradient~\cite{Stevens1997SIAM-JAM,Painter2002CanApplMathQuart,Hillen2009JMathBiol,Bubba2019JTheorBiol}. 
We assume that each agent follows a cue-dependent movement policy obtained by replacing $\beta V^*_\mu$ in the optimal movement policy~\eqref{eq:EMDP-greedy-policy} with the endogenous cue concentration $Z_t=\beta V_t$:
\begin{equation}
    \label{eq:dist-G-learning-policy}
    \pi[Z_t](s'|s)\coloneqq \frac{p(s'|s)\exp(\gamma Z_t(s'))}{\sum_{s'\in\mathcal{S}}p(s'|s)\exp(\gamma Z_t(s'))},\quad s,s'\in\mathcal{S}.    
\end{equation}
\DraftNote{[C8]}
For the current population density $\mu_t$, the cue-guided movement policy $\pi[Z_t]$ is approximately optimal if $Z_t$ closely approximates $\beta V^*_{\mu_t}$.
When all agents follow this policy based on the shared endogenous cue $Z_t$, the population density $\mu_t$ evolves according to
\begin{equation}
    \label{eq:mean-field-evolution}
    \mu_{t+1}(s)=\sum_{s'\in\mathcal{S}}\mu_t(s')\pi[Z_t](s|s'),\quad s\in\mathcal{S}.    
\end{equation}

\DraftNote{[C8]}
While moving, agents also modulate the endogenous cue $Z_t$ using their local observations.
Here we introduce simple, biologically plausible cue dynamics to improve the estimate $V_t=Z_t/\beta$ of $V^*_{\mu_t}$ and thereby improve agents' cue-guided movement.
With cue diffusion included, the dynamics are given by (see SM Sec.~\ref{sec:algorithm-derivation} for a derivation): 
\begin{subequations}\label{eq:dist-G-learnSMFG-diff}
    \begin{alignat}{3}
        Z_{t+1}(s)&-Z_{t}(s)=&&\mu_t(s)\alpha\Delta Z_{t}(s)
        \notag\\
        &&-&D \sum_{s'\in \mathcal{S}}L_{ss'} Z_{t}(s'),\quad s\in\mathcal{S},\\
        \Delta Z_t(s)&\coloneqq -Z_{t}(s)&+&\beta r_g(s;\mu_t)\notag\\
        &&+&\ln\sum_{s'\in\mathcal{S}}p(s'|s)\exp(\gamma Z_{t}(s')).
        \end{alignat}
\end{subequations}
where $L$ is the graph Laplacian of the environment $\mathcal{G}$ and $\alpha\in(0,1)$ represents the rate of local cue modulation by agents.
In Eq.~\eqref{eq:dist-G-learnSMFG-diff}, each agent modulates $Z_{t}(s)$ by $\alpha \Delta Z_t(s)$ based on nearby endogenous cue $Z_t(s')$ and immediate reward $r_g(s;\mu_t)$, which depends on the local exogenous guiding signal and population density.
The cue also diffuses between neighboring vertices at a constant rate $D$, independently of the local population density.

\DraftNote{[C9]}
If the coupled dynamics~\eqref{eq:mean-field-evolution} and \eqref{eq:dist-G-learnSMFG-diff} become nearly stationary, the population density $\mu_t$ and associated movement policy $\pi[Z_t]$ approximately satisfy the equilibrium condition~\eqref{eq:SMFG-NE}, provided that cue diffusion is sufficiently small compared to local cue modulation (see Prop.~\ref{prop:proposed-near-stationarity} in SM Sec.~\ref{sec:proposed-stationarity}).
Population consistency~\eqref{eq:SMFG-NE-stationarity} immediately follows from the stationarity of density under the movement policy $\pi[Z_t]$~\eqref{eq:dist-G-learning-policy}.
Individual optimality~\eqref{eq:SMFG-NE-optimality} is approximately satisfied because a stationary cue field $Z_t$ under Eq.~\eqref{eq:dist-G-learnSMFG-diff} indicates that the field closely approximates the optimal value $\beta V^*_{\mu_t}$ for the current population density $\mu_t$.
These learning dynamics are a distributed~\cite{Mnih2016PMLR}, sequentially regularized version of coupled learning algorithms for the SMFG problems~\cite{Subramanian2019AAMAS,Angiuli2022MCSS}.

\subsection{Continuous limit of coupled dynamics}
\DraftNote{[C10]}
We next derive a continuous limit of the coupled dynamics~\eqref{eq:mean-field-evolution} and \eqref{eq:dist-G-learnSMFG-diff} and show its structural correspondence to well-established Keller-Segel type models for chemotactic aggregation~\cite{Keller1970JTheorBiol,Tsimring1995PRL,Woodward1995BiophysJ,Brenner1998BiophysJ,Tyson1999ProcRSocLondB,Painter2019JTB}.
This extends the correspondence established in our previous work on accumulation near sources of exogenous guiding signals~\cite{Kato2025PRXLife} to density-dependent aggregation.
\DraftNote{[C10]}
For simplicity, we derive the limit on a one-dimensional lattice $\mathcal{S}=\{0,\pm h,\pm 2h,\ldots\}$.
Consider the space-time scale $\Delta t=h^2$, and let the intrinsic movement policy $p^h_{\Delta t}$ be a lazy random walk: $p^h(x|x)=1-\bar{\sigma}\Delta t/h^2,\ p^h(x\pm h|x)=\bar{\sigma}\Delta t/(2h^2)$.
We scale the rate of local cue modulation, cue diffusion rate, and local reward as follows: $\alpha^h_{\Delta t}=\bar{\alpha}\Delta t/h$, $D^h_{\Delta t}=\bar{D}\Delta t/h^2$, $r^h_{g}(s;\mu)=\bar{r}_g(s;\mu(s)/h)$.
As $h\to 0$, the coupled dynamics yield the following partial differential equations (see SM Sec.~\ref{sec:continuous-limit} for a derivation in a general setting with a density-dependent intrinsic movement policy):
\begin{subequations}               
    \label{eq:lin-continuous}
    \begin{align}
        \pt_t\mu&=-\nabla\cdot\left(\mu\gamma \bar{\sigma} \nabla Z\right)+\frac{\bar{\sigma}}{2}\nabla^2\mu,\label{lin-continuous-popu}\\
        \pt_t Z&=-\left[\bar{\alpha} (1-\gamma)\mu Z\right]+\left[\bar{\alpha}\beta\mu \bar{r}_{g}(\cdot;\mu(\cdot))\right]+\bar{D}\nabla^2 Z,\label{lin-continuous-chem}
    \end{align}
\end{subequations}
where $\mu(t,x)\in\mathbb{R}_{\geq 0}$, $Z(t,x)\in\mathbb{R}_{\geq 0}$, $g(x)\in\mathbb{R}_{\ge0}$ are the population density, endogenous cue concentration, and exogenous guiding signal concentration, respectively. 

\DraftNote{[C10]}
The equations~\eqref{eq:lin-continuous} obtained in the continuous limit have the same structure as the Keller-Segel type models of microbial aggregation mediated by chemoattractants without growth~\cite{Keller1970JTheorBiol,Tsimring1995PRL,Woodward1995BiophysJ,Brenner1998BiophysJ,Tyson1999ProcRSocLondB}:
\begin{subequations}
    \label{eq:KS}
    \begin{align}
        \pt_t n&=-\nabla \cdot \left(n\chi\nabla f(c)\right)+D_n\nabla ^2n,\label{eq:KS-popu}\\
        \pt_t c&=-\left[\eta nc\right] + \left[np(n,s)\right]+D_s\nabla^2 c\label{eq:KS-chem},
    \end{align}
\end{subequations}
where $n(t,x)$, $c(t,x)$, $s(x)$, $f(c)$, and $p(n,s)$ denote the microbial population density, the attractant concentration, the nutrient concentration, the cue-response function, and the attractant production rate per cell, respectively. 
$\chi$, $D_n$, $D_s$, and $\eta$ represent the chemotactic coefficient, the microbial diffusion coefficient, the diffusion coefficient of the attractant, and the attractant uptake rate by bacteria, respectively.
The population equations in Eqs.~\eqref{lin-continuous-popu} and \eqref{eq:KS-popu} have the same structure for the linear response $f(c)=c$.
Both cue equations~\eqref{lin-continuous-chem} and \eqref{eq:KS-chem} contain a density-dependent degradation term proportional to the cue concentration:  $-[\bar{\alpha}(1-\gamma)\mu Z]$ and $-\eta nc$.
Both production terms depend on local population density and exogenous guiding signal concentration when the reward $r_g(s;\mu(s))$ depends only on these quantities: $[\bar{\alpha}\beta\mu \bar{r}_g(\cdot;\mu(\cdot))]$ and $[np(n,s)]$. 
\DraftNote{[C10]}
This correspondence 
suggests that chemotactic aggregation can be associated with equilibrium learning by decentralized agents with density-dependent incentives.

\DraftNote{[C2][C10]}
A pioneering study by Pezzotta \textit{et al.}~\cite{Pezzotta2018PRE} identified a link between the Keller-Segel equation and mean field game equations, interpreting chemotaxis in a stationary endogenous cue field as movement along the optimal state value gradient.   
Our previous work~\cite{Kato2025PRXLife} employed reinforcement learning (RL) theory to derive the adaptation process toward optimal movement, identifying an optimal coupling between chemotactic movement and cue modulation.
However, that framework addressed only accumulation near sources of exogenous guiding signals with density-independent rewards.
\DraftNote{[C8]}
The formulation developed here incorporates density-dependent rewards into these learning dynamics, describing adaptation toward equilibrium patterns.
This extension encompasses not only self-organized aggregation without exogenous guiding signals but also spatial pattern formation driven jointly by exogenous guiding signals and density-dependent incentives.

%% file: 3-experiment.tex
\section{Numerical experiments}
\label{sec:numexp}
\DraftNote{[Aim and context of numerical experiments]}
\DraftNote{[C11]}
The derived coupled dynamics describe adaptation toward equilibrium patterns, in which individuals optimize their own movement.
The population-level efficiency of these patterns, however, remains to be examined.
In this section, we consider self-organized aggregation and use numerical simulations to demonstrate that the dynamics can form equilibrium patterns and to evaluate their population-level efficiency.
Self-organized aggregation mediated by extracellular communication has been observed in various biological systems ranging from bacteria~\cite{Budrene1991Nature,Park2003PNAS,Scheidweiler2024NatCommun}, social amoebae~\cite{Park2004EMBOJ,Bonner2009} to developing tissues~\cite{Ho2019PLoSBiol} and tumor cells~\cite{Puliafito2015SciRep,Bubba2019JTheorBiol}. 
Aggregation is also of theoretical interest.
Unlike dispersal, which yields a unique dispersed equilibrium pattern (see SM Prop.~\ref{prop:SMFG-monotone-uniqueness}), aggregation may admit multiple equilibrium patterns.
Local aggregates with different density profiles can be sustained in different spatial configurations by agents that cannot improve their own outcomes by unilaterally changing their movement.
This multiplicity raises a further question about which equilibrium pattern is realized.
In pattern selection, the relative timescales of movement and cue modulation are another key factor alongside their coupling established in our previous work~\cite{Kato2025PRXLife}.
These timescales can change the stability and basins of attraction of patterns, thereby affecting pattern selection and population-level efficiency.

In Sec.~\ref{sec:fast-modulation}, we first demonstrate that the dynamics with fast cue modulation form multiple equilibrium patterns, some of which are comparable to the population-level optimum.
In Sec.~\ref{sec:slow-modulation}, we show that slower cue modulation can form more coherent patterns with higher population-level efficiency.

\subsection{Self-organized aggregation often yields equilibrium patterns with different efficiencies}
\label{sec:fast-modulation}
\DraftNote{(Aim of this section: To demonstrate that rational agents self-organize socially non-optimal patterns due to the locality of their aggregating incentive.)}

\DraftNote{[Introduction to bacterial self-organized aggregation: Phenomenon, mapping to the problem, we will use the problem]}
\DraftNote{[C11]}
Biological systems often self-organize spatial patterns in complex environments.
Particularly, bacterial populations can form aggregates depending on the surrounding topology even in the absence of exogenous guiding signals~\cite{Park2003PNAS,Scheidweiler2024NatCommun}.
We consider self-organized aggregation in an environment with branching structures, which may make it difficult for spatially scattered agents to coordinate. 

\begin{figure*}[tbp]
    \hfill
    \begin{minipage}[b]{\linewidth}
        \centering
        \includegraphics[width=\textwidth]{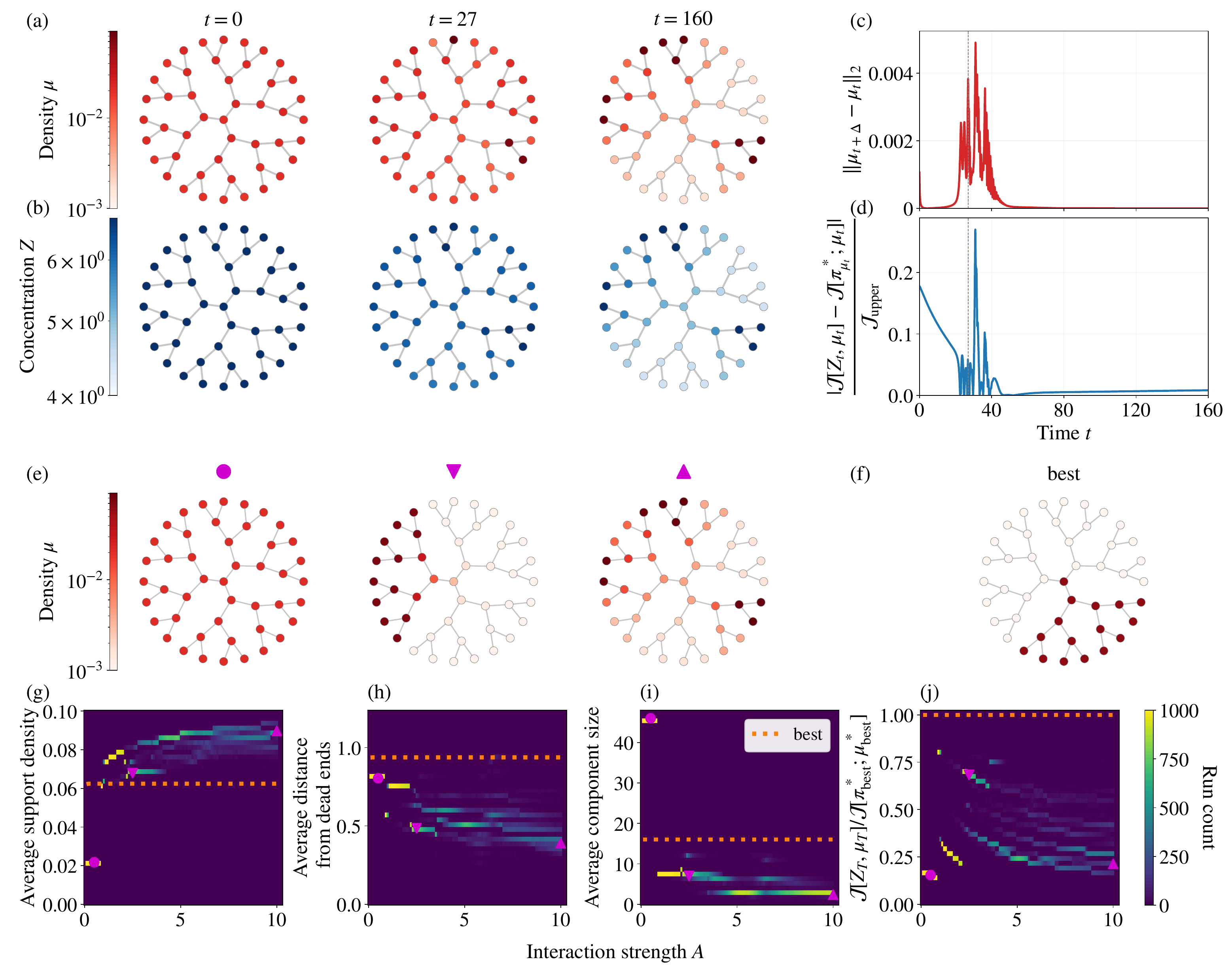}
    \end{minipage}
    \hfill
    \caption{
    Numerical simulations of self-organized aggregation with cue modulation faster than movement.
    (a, b) Snapshots at $t=0,27,$ and $160$ for one trial of self-organized aggregation at $A=10$ with random seed 278, using movement~\eqref{eq:mean-field-evolution} and fast cue modulation~\eqref{eq:dist-G-learnSMFG-diff}.
    Vertex colors on the tree graph indicate population density $\mu_t$ in (a) and endogenous cue concentration $Z_t$ in (b).
    (c) Time evolution of density change over successive saved epochs $\|\mu_{t+\Delta}-\mu_t\|_2$, with $\Delta=0.05$. 
    (d) Normalized gap $|\mathcal{J}[Z_t,\mu_t]-\mathcal{J}[\pi^*_{\mu_t};\mu_t]|/\mathcal{J}_\mathrm{upper}$ between the objective estimate and the optimal objective for the current density.
    The normalization term is a uniform upper bound on the objective: $\mathcal{J}_\mathrm{upper}\coloneqq \max_{\mu\in\Delta(\mathcal{S})}\sum_{s\in\mathcal{S}}\mu(s)r_g(s;\mu(s))/(1-\gamma)$. 
    (e) Final population density profiles $\mu_T$ from selected trials at different interaction strengths $A$: $(A,\textrm{seed})=(0.5,1),(4,108),(10,278)$ for $\bullet,\blacktriangledown,\blacktriangle$, respectively.
    We assess stationarity at selected epochs and stop the simulation at $T$ when the density change and normalized cue change between successive assessments fall below prescribed thresholds $\epsilon^\mu$ and $\epsilon^Z$, respectively.
    When the uniform state is unstable, we additionally require the $L^1$ distance from uniform to exceed $\delta_\mathrm{uni}$.
    (f) The density profile $\mu^*_\mathrm{best}$ with the highest population-level objective value $\mathcal{J}[\pi;\mu^\pi]$ found numerically for each $A$. 
    The same profile was obtained for all tested values of $A$.
    (g-j) Histograms of final-pattern metrics across trials at different values of $A$. 
    Colors indicate the number of trials in each bin.
    Magenta markers indicate the patterns shown in (e), and orange dotted lines indicate the best pattern shown in (f).
    (g) Mean population density across vertices whose final density is at least 99.8\% of the uniform value $1/|\mathcal{S}|$.
    (h) Population-averaged distance to the nearest dead-end vertex.
    (i) Population-weighted mean size of connected clusters formed by the vertices selected in (g).
    (j) Population-level efficiency relative to the best pattern found numerically, measured by $\mathcal{J}[Z_T,\mu_T]/\mathcal{J}[\pi^*_\mathrm{best};\mu^*_\mathrm{best}]$.    
    The parameter values used are $\beta=1.0,\gamma=0.8,\varepsilon=0.05,\alpha=0.5,D=0.0125$, $Z_0=10.0,\delta=0.1$, $a=2,b=10.0,K=0.072$ and $\epsilon^\mu=10^{-5},\epsilon^Z=5\times10^{-5},\delta_\mathrm{uni}=0.2$.
    We use 1000 trials for each $A\in[0.1,10]$.
    }
    \label{fig:fast-modulation}
\end{figure*}

\DraftNote{[Experimental setup]}
\DraftNote{[C11]}
Specifically, we consider a simple tree environment of depth four with two children at each vertex (Fig.~\ref{fig:fast-modulation}).
In the following numerical experiments, we assume a lazy Laplacian random walk for the intrinsic movement policy $p$ of agents: $p(s|s) = 1-\varepsilon |\mathcal{N}(s)|$ and $p(s'|s) = \varepsilon,\ s' \in \mathcal{N}(s)$, where the parameter $\varepsilon$ represents the movement speed.
The stationary distribution of this intrinsic random walk $p$ is uniform.
\DraftNote{[C11]}
The exogenous guiding signal $g$ is assumed to be uniform across the graph, representing that there is no explicit target signal available to the agents.
We represent the incentive to aggregate by a local reward adapted from a phenomenological model~\cite{Tyson1999ProcRSocLondB}: $r_g(s;\mu(s))=A \frac{(\mu(s)/K)^a}{1+(\mu(s)/K)^b},\ 0<a<b$ with a maximum at a density $K(a/(b-a))^{1/b}$.
Here, $A$ sets the interaction strength, $K$ sets the density scale, and $a$ and $b$ determine the shape of the reward function.
This function represents the benefit of aggregation at low density, such as cooperative metabolism, and the costs of crowding at high density, such as resource depletion.
\DraftNote{[C11]}
We consider initial states with a uniform cue concentration $Z(s)= Z_0$ and a randomly perturbed uniform density $\mu_0(s)\propto 1+\xi(s)$ for each $s\in\mathcal{S}$.
Here, $\xi(s)\sim\operatorname{Uni}[-\delta,\delta]$ with a perturbation strength $\delta\in(0,1)$.
Then, agents collectively modify the endogenous cue concentration to explore the environment and form aggregates (Fig.~\ref{fig:fast-modulation}).
We first examine a regime in which cue modulation is faster than movement, i.e. $\alpha>\varepsilon$.

\DraftNote{[Result 1.1 (Trial-wise observation, Relevance to real experiments): aggregation among dead-ends]}
\DraftNote{[C12]}
We will see through a single trial that the population exhibits aggregation qualitatively consistent with previous experimental observations.
Although the initially uniform cue field provides no directional guidance, the dispersed agents form aggregates near dead-end vertices through local responses (Fig.~\ref{fig:fast-modulation} (a, b), $t=0,27$).
The population eventually reaches a stationary state, with its aggregates remaining scattered among these vertices (Fig.~\ref{fig:fast-modulation} (a, b), $t=160$).
Similar accumulation at dead ends has been observed in bacterial populations~\cite{Park2003PNAS,Scheidweiler2024NatCommun}.

\DraftNote{[Result 1.2 (Convergence to NE)]}
This aggregation process can also be viewed as adaptation toward an equilibrium pattern~\eqref{eq:SMFG-NE}.
During this adaptation, the individual movement policy is expected to remain approximately optimal for the current density (Eq.~\eqref{eq:SMFG-NE-optimality}) because fast cue modulation keeps $Z_t$ close to $\beta V^*_{\mu_t}$ as the population redistributes.
Consistent with this expectation, an estimate of the objective $\mathcal{J}[Z_t,\mu_t]$ remains close to the optimal objective $\mathcal{J}[\pi^*_{\mu_t};\mu_t]$ for the current density throughout the trial (Fig.~\ref{fig:fast-modulation} (d)).
Here, $\mathcal{J}[Z_t,\mu_t]\coloneqq \frac{1}{\beta}\sum_{s\in\mathcal{S}}\mu_t(s)Z_t(s)$ and $\mathcal{J}[\pi^*_{\mu_t};\mu_t]=\max_{\pi}\sum_{s\in\mathcal{S}}\mu_t(s)V^\pi_{\mu_t}(s)$.
$\mathcal{J}[\pi^*_{\mu_t};\mu_t]$ is computed by value iteration for each $\mu_t$ (see SM Sec.~\ref{sec:ermdp-foundation-draft}).
The density change also eventually becomes small, indicating that the final pattern is close to an equilibrium pattern (Fig.~\ref{fig:fast-modulation} (c, d)).

\DraftNote{[Result 2.0 (Population-level inefficiency of equilibrium patterns)]}
\DraftNote{[C13]}
Individual movement in an equilibrium pattern is nearly optimal, as indicated by the objective estimate $\mathcal J[Z_t,\mu_t]$, but the pattern may still be inefficient at the population level.
To highlight its population-level efficiency, we compare its objective estimate with the highest population-level objective $\mathcal{J}[\pi;\mu^\pi]$ found numerically (see SM Sec.~\ref{sec:stationary-mfc-numerics} for details).
We denote the movement policy of the best pattern by $\pi^*_{\mathrm{best}}$ and its stationary density by $\mu^*_{\mathrm{best}}=\mu^{\pi^*_{\mathrm{best}}}$.
We quantify the population-level efficiency of equilibrium patterns relative to the best pattern using the ratio $\mathcal{J}[Z_T,\mu_T]/\mathcal{J}[\pi^*_{\mathrm{best}};\mu^*_\mathrm{best}]$.
When it is close to 1, the pattern is comparable to the best pattern, which sometimes requires centralized control.
Its minimum among equilibria can be associated with the well-known game-theoretic metric called the Price of Anarchy~\cite{Roughgarden2002JACM}, which measures the suboptimality caused by decentralized behaviors.

\DraftNote{[Result 2.1 (Most the final patterns are not optimal)]}
\DraftNote{[C13]}
Across multiple trials with different interaction strengths and initial density perturbations, we observe various final patterns (Fig.~\ref{fig:fast-modulation} (e)). 
Many of these patterns, however, fall short of the best pattern (Fig.~\ref{fig:fast-modulation} (f)) in terms of population-level efficiency.
When the interaction strength $A$ is small, the population becomes uniformly distributed (Fig.~\ref{fig:fast-modulation} (e) $\bullet$ and (g)) and its population-level efficiency is low (Fig.~\ref{fig:fast-modulation} (j)).
When $A$ is moderate, the population forms a few clusters near dead-end vertices (Fig.~\ref{fig:fast-modulation} (g-i)), with relative population-level efficiency exceeding 60\% in some trials (Fig.~\ref{fig:fast-modulation} (e) $\blacktriangledown$ and (j)).
However, as $A$ increases further, the population is more fragmented across dead-end vertices (Fig.~\ref{fig:fast-modulation} (e) $\blacktriangle$ and (g-i)), while its population-level efficiency fails to improve and can even decline (Fig.~\ref{fig:fast-modulation} (j)).
These results suggest that fast cue modulation can yield patterns comparable to the best pattern at moderate interaction strengths, whereas stronger interactions often lead to fragmented, inefficient patterns.

\DraftNote{[Implications]}
\DraftNote{[C13]}
Our theory allows us to interpret self-organized chemotactic aggregation as an adaptation process toward equilibrium patterns through coupled movement and cue modulation.
Aggregation can thus arise from agents improving their own movement under a shared local density-dependent incentive, not merely from the instability of a homogeneous pattern~\cite{Hillen2009JMathBiol}.
However, the local incentive can strengthen aggregation without promoting coordination across the population. 
Under fast cue modulation, cue concentrations can rapidly rise around local aggregates, attracting more agents and reinforcing inefficient patterns of spatially separated aggregates.
A stronger incentive to aggregate can even worsen fragmentation and reduce population-level efficiency, with a similar loss of efficiency reported in a congestion avoidance problem~\cite{Lauriere2021ProcSympApplMath}. 

\subsection{Slow cue modulation can yield more efficient patterns}
\label{sec:slow-modulation}

\DraftNote{(Aim of this section: To demonstrate that slow modulation dynamics reach clustered patterns with slightly better objective than fast modulation dynamics.)}

\DraftNote{[Introduction to a timescale problem: mapping to the problem]}
\DraftNote{[C14][C15]}
Populations with the same incentive may form different patterns depending on the relative timescales of cue modulation and movement.
A link between cell motility and aggregation morphology has been observed in social amoebae~\cite{Park2004EMBOJ}, where slower-moving mutant cells show delayed aggregation and subsequent fragmentation.
The preceding results in Sec.~\ref{sec:fast-modulation} suggest that fast cue modulation can also yield spatially fragmented, inefficient patterns by reinforcing local aggregation too quickly.
Slower cue modulation may instead allow the population to redistribute before the cue reinforces local aggregation, enabling more coordinated and efficient aggregation.
\DraftNote{[C14]}
To test this hypothesis, we use the same setup as in Sec.~\ref{sec:fast-modulation}, but with cue modulation slower than movement $\alpha<\varepsilon$, while keeping $\varepsilon$ and $D/\alpha$ fixed.

\begin{figure*}[tbp]
    \hfill
    \begin{minipage}[b]{\linewidth}
        \centering
        \includegraphics[width=\textwidth]{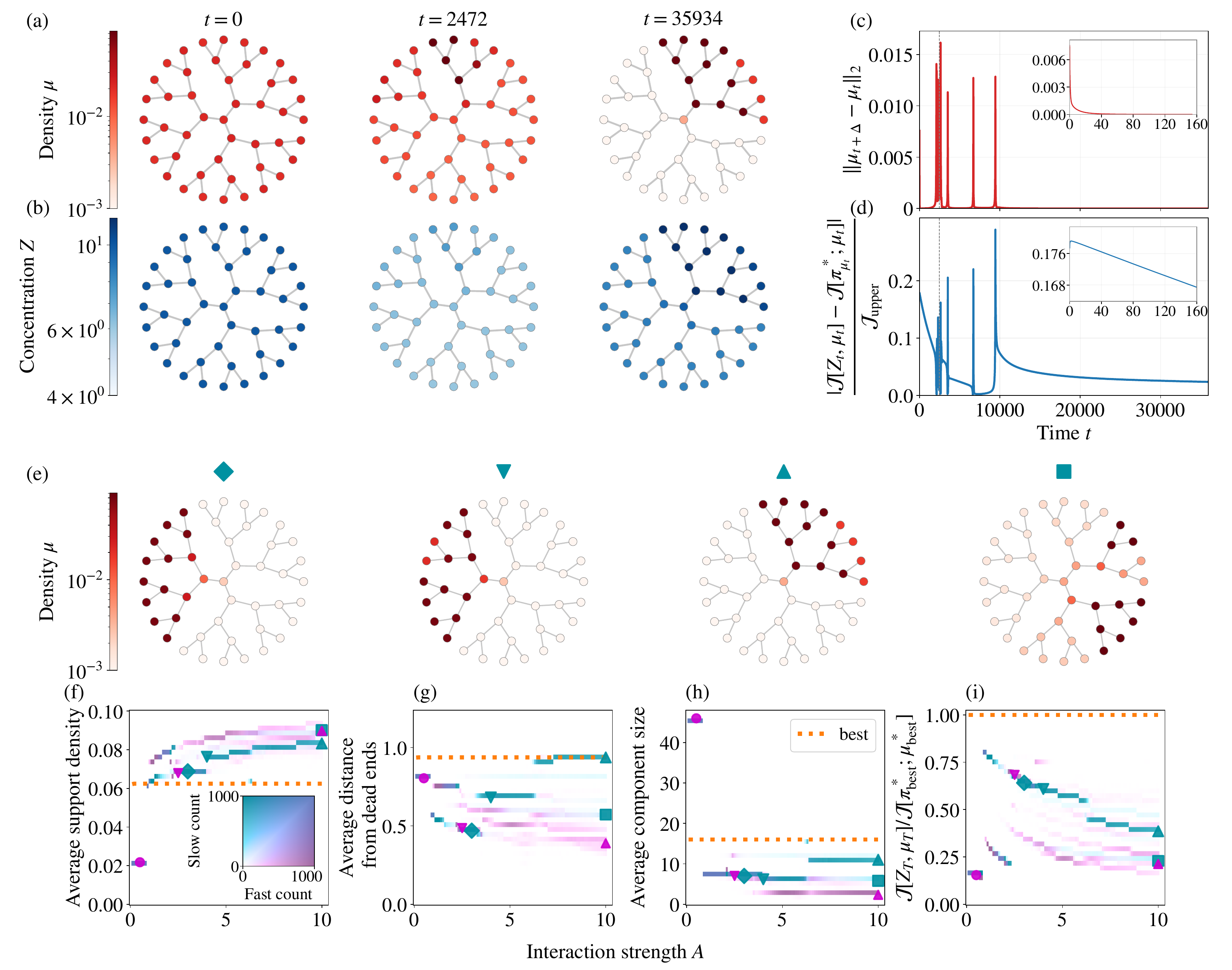}
    \end{minipage}
    \hfill
    \caption{
    Numerical simulations of self-organized aggregation with cue modulation slower than movement, including comparisons with fast cue modulation.
    (a, b) Snapshots at $t=0,2472,$ and $35934$ for one trial of self-organized aggregation at $A=10$ with random seed 278, using movement~\eqref{eq:mean-field-evolution} and slow cue modulation~\eqref{eq:dist-G-learnSMFG-diff}.
    Vertex colors on the tree graph indicate population density $\mu_t$ in (a) and endogenous cue concentration $Z_t$ in (b).
    (c) Time evolution of density change over successive saved epochs $\|\mu_{t+\Delta}-\mu_t\|_2$, with $\Delta=6$. 
    (d) Normalized gap $|\mathcal{J}[Z_t,\mu_t]-\mathcal{J}[\pi^*_{\mu_t};\mu_t]|/\mathcal{J}_\mathrm{upper}$ between the objective estimate and the optimal objective for the current density.
    The normalization term is a uniform upper bound on the objective: $\mathcal{J}_\mathrm{upper}\coloneqq \max_{\mu\in\Delta(\mathcal{S})}\sum_{s\in\mathcal{S}}\mu(s)r_g(s;\mu(s))/(1-\gamma)$.
    (e) Final population density profiles $\mu_T$ from selected trials at different interaction strengths $A$: $(A,\textrm{seed})=(3,1),(4,1),(10,278),(10,2)$ for $\rotatedsquare,\blacktriangledown,\blacktriangle,\blacksquare$, respectively.
    We assess stationarity at selected epochs and stop the simulation at $T$ when the density change and normalized cue change between successive assessments fall below prescribed thresholds $\epsilon^\mu$ and $\epsilon^Z$, respectively.
    When the uniform state is unstable, we additionally require the $L^1$ distance from uniform to exceed $\delta_\mathrm{uni}$.
    (f-i) Histograms of final-pattern metrics across trials at different values of $A$, comparing slow and fast cue modulation.
    Colors indicate the numbers of slow- and fast-modulation trials in each bin.
    Teal and magenta markers indicate the patterns shown in (e) and Fig.~\ref{fig:fast-modulation} (e), respectively.
    Orange dotted lines indicate the best pattern shown in Fig.~\ref{fig:fast-modulation} (f).
    (f) Mean density across vertices whose final density is at least 99.8\% of the uniform value $1/|\mathcal{S}|$. 
    (g) Population-averaged distance to the nearest dead-end vertex.
    (h) Population-weighted mean size of connected clusters formed by the vertices selected in (f).
    (i) Population-level efficiency relative to the best pattern found numerically, measured by $\mathcal{J}[Z_T,\mu_T]/\mathcal{J}[\pi^*_\mathrm{best};\mu^*_\mathrm{best}]$.
    The parameter values used are $\beta=1.0,\gamma=0.8,\varepsilon=0.05,\alpha=0.005,D=0.000125$, $Z_0=10.0,\delta=0.1$, $a=2,b=10.0,K=0.072$ and $\epsilon^\mu=10^{-5},\epsilon^Z=5\times10^{-5},\delta_\mathrm{uni}=0.2$.
    We use 1000 trials for each $A\in[0.1,10]$.
    }
    \label{fig:slow-fast}
\end{figure*}

\DraftNote{[Trial-wise observation: Cluster formation and convergence to NE]}
\DraftNote{[C14]}
In a single trial under slower cue modulation (Fig.~\ref{fig:slow-fast} (a-d)) from the same initial state as in Fig.~\ref{fig:fast-modulation} (a-d), the population forms a connected aggregate and approaches an equilibrium pattern~\eqref{eq:SMFG-NE} with higher population-level efficiency.
The initially perturbed population density first becomes nearly uniform and remains approximately stationary for a time.
As the cue field evolves, a cluster forms and grows by drawing in agents from the rest of the population (Fig.~\ref{fig:slow-fast} (a-d) at $t=2472$).
The population eventually forms a stationary, connected aggregate.
The objective estimate $\mathcal{J}[Z_T,\mu_T]$ is then close to the optimal objective $\mathcal{J}[\pi^*_{\mu_T};\mu_T]$ for the final density, indicating that the pattern is close to an equilibrium pattern~\eqref{eq:SMFG-NE} (Fig.~\ref{fig:slow-fast} (a-d) at $t=35934$).
This more coherent aggregation with faster movement relative to cue modulation is qualitatively consistent with observations in \textit{Dictyostelium}~\cite{Park2004EMBOJ}.

\DraftNote{[Many trial observation: Clusters and their superiority at intermediate interaction strength]}
\DraftNote{[C14]}
Across trials, slow modulation yields more coherent final patterns with higher population-level efficiency than those under fast modulation, especially at moderate interaction strengths.
For each $A=3,4$, the population forms several aggregates near different dead-end vertices and often achieves higher population-level efficiency than under fast modulation (Fig.~\ref{fig:slow-fast} (e) $\rotatedsquare,\blacktriangledown$ and (f-i)).
At stronger interaction strengths, the population may form a single connected cluster farther from dead ends, but its population-level efficiency is only slightly higher than under fast modulation (Fig.~\ref{fig:slow-fast} (e) $\blacktriangle,\blacksquare$) (Fig.~\ref{fig:slow-fast} (f-h)).

\DraftNote{[Interpretation and Implications]}
\DraftNote{[C15]}
Slower cue modulation may reduce excessive local accumulation and improve population-level efficiency at moderate interaction strengths, although the population takes longer to approach an equilibrium pattern.
Empirical findings in mean field learning likewise suggest that relative timescales can change basins of attraction, with faster population dynamics favoring equilibria closer to the social optimum~\cite{Angiuli2022MCSS,An2024arXiv}.
These results suggest that slower cue modulation relative to movement could confer an evolutionary advantage by promoting more efficient aggregation.

%% file: 4-conclusion-discussion.tex
\section{Summary and Discussion}
\label{sec:conclusion}
\DraftNote{[How we formulate and model for what aim and what directly follows]}
\DraftNote{[C2][C5]}
In this study, we have developed a normative framework for evaluating efficiency of spatial patterns formed by chemotactic cells.
We have characterized efficient spatial patterns achievable by decentralized cell populations as Nash equilibria of a stationary mean field game.
At equilibrium, each cell optimizes its own movement for its own outcome given the population density generated by the other cells following the same strategy. 
\DraftNote{[C8][C10]}
We have derived coupled dynamics of movement and cue modulation: agents cooperatively modulate an endogenous cue field based on local observations to improve their own movement for the current population density.
The derived dynamics approach equilibrium patterns and, as in our previous work, structurally correspond to phenomenological Keller-Segel type models.
\DraftNote{[C11][C12][C13]}
We have applied our framework to self-organized aggregation in complex environments  
and numerically demonstrated that the dynamics can form multiple equilibrium patterns. 
Some of these patterns attain population-level efficiencies comparable to the optimum under moderate incentives to aggregate. 
\DraftNote{[C14][C15]}
Our results have further shown that the relative timescales of movement and cue modulation affect pattern selection and population-level efficiency. 
Slower cue modulation can yield more coherent and efficient patterns, possibly by allowing redistribution before cue-mediated feedback reinforces local accumulation.
These findings reveal fundamental constraints on collective optimization in decentralized cell populations and suggest a functional role for relative timescales in coordination through environmental feedback.

\subsection{Generality of the formulation}
\DraftNote{[C16]}
Our formulation encompasses a broader class of dynamics and spatial patterning problems than those examined here.
As in our previous work~\cite{Kato2025PRXLife}, we can derive various dynamics with nonlinear gradient responses by replacing the linear mapping between the state value estimate $V_t$ and endogenous cue concentration $Z_t$ with a nonlinear mapping.
The framework also accommodates density-dependent accumulation in heterogeneous exogenous guiding signal fields, as in immune cells accumulating around injury sites while avoiding excessive inflammation~\cite{Glaser2021CurrOpiCellBiol}.


Our formulation can be extended to density-dependent intrinsic movement $p(s'|s;\mu)$ (See SM Sec.~\ref{sec:entropy-SMFG}).
Density-dependent mechanical interactions indeed occur in migrating cell populations, including crowding~\cite{Sadati2013Diff,Colin2019NatCommun}, cell adhesion~\cite{Friedl2017CSHPerspectBiol,Laganenka2016NatComm}, and cell--cell repulsion~\cite{Stramer2017NatRevMolCellBiol}.
These interactions can be represented phenomenologically by random walks that depend on population density~\cite{Painter2002CanApplMathQuart,Anguige2009JMathBiol,Bubba2019JTheorBiol,Buttenschon2018JMathBiol}.
With a density-dependent intrinsic movement policy, the stationary distribution need not be unique even under a uniform endogenous cue field, potentially allowing a wider variety of equilibrium patterns.
It would be interesting to examine how relative timescales affect pattern selection and population-level efficiency when density directly affects intrinsic movement as well as cue modulation.

\subsection{Extension to multiple populations and nonstationary patterning}
\DraftNote{[C16]}
This study has focused on stationary spatial patterns formed by a single cell type.
However, spatial patterning in development, immune cell migration, and cancer metastasis often involves interactions among functionally heterogeneous cell populations~\cite{Mayor2016NatRevMolCellBiol,Shellard2020PhilTransRSocB,Alanko2023SciImm}.
Interactions between different cell types are often asymmetric and can give rise to not only stationary spatial organization such as spatial coexistence or segregation but also to autonomous nonstationary patterns~\cite{Painter2019JTB}. 
During nervous system development, for example, sustained collective migration emerges through chase-and-run interactions: neural crest cells chemotax toward signals released by placodal cells, while placodal cells move away upon contact with neural crest cells~\cite{theveneau2013NatCellBiol}.
Examining how timescales influence the collective performance of such sustained patterning would require extending our framework to heterogeneous populations~\cite{Subramanian2020AAMAS} and developing an equilibrium formulation for sustained patterning.
Furthermore, cellular phenotypes can also change in response to local conditions~\cite{Mclennan2015DevBiol}.
Incorporating phenotypic switching alongside migration~\cite{Macfarlane2022BullMathBiol,Painter2023JApplMath} would allow us to examine how cells should switch between phenotypes to support division of labor during spatial patterning.

\subsection{Functional role of relative timescales of intercellular communication and movement}

\DraftNote{[C15]}
The dependence of the pattern efficiency on the relative timescales may offer a functional perspective on why these timescales vary among cell populations.
Bacteria and social amoebae gradually form patterns on timescales longer than those of movement while modulating chemical cue fields~\cite{Laganenka2016NatComm,Park2004EMBOJ}, suggesting that cue modulation might be slower than movement.
In contrast, immune and cancer cells modulate chemical cue fields on timescales shorter than those of movement~\cite{Tweedy2020Science,Strickland2024DevCell}, providing directional guidance before substantial population redistribution.
These differences may reflect the functional demands placed on cell populations in different environments~\cite{Clark2005PNAS}.
When aggregation provides lasting benefits, such as antimicrobial tolerance in bacteria~\cite{Laganenka2016NatComm}, the performance of the stationary pattern may contribute more to fitness than the speed of pattern formation.
Conversely, when rapid accumulation is critical, as in immune responses to proliferating pathogens~\cite{Hopke2020NatCom}, speed may take precedence over the performance of the final pattern, potentially favoring faster cue modulation.
A promising direction is to extend our framework to evaluate performance during pattern formation and investigate possible trade-offs between aggregation speed and the population-level efficiency of the final pattern.

\subsection{Efficiency maximization near the transition}
Under both fast and slow cue modulation, the relative population-level efficiency exhibits a discontinuous jump to a maximum, followed by a monotonic decline with increasing interaction strength $A$ (Figs.~\ref{fig:fast-modulation}(j) and \ref{fig:slow-fast}(i)).
The jump corresponds to a bifurcation from a uniform equilibrium pattern to a non-uniform aggregated equilibrium pattern (See SM Sec.~\ref{sec:uniform-graph-instability}).
Such performance maximization near the bifurcation echoes the longstanding criticality hypothesis in nonlinear dynamics and statistical physics: self-organized systems operating near criticality can gain functional benefits, such as enhanced susceptibility to external perturbations and increased memory capacity~\cite{Munoz2018RevModPhys}.
However, the functional significance of criticality remains controversial, highlighting the need for a theoretical framework that unifies functional optimality and the underlying dynamics.
Our study shows that maximized population-level efficiency near criticality emerges in collective dynamics derived from optimization in decentralized populations.
Such dynamical models derived from optimization principles~\cite{Tottori2025PhysRevRes,Tottori2026PhysRevLett} can provide a solid understanding of the relationship between functional optimality and emergent dynamical phenomena.

%% file: AppendixA-Theory.tex
\label{sec:MDPs}
This section provides the theoretical basis for our formulation and mathematical analysis of the coupled dynamics in the main text.
Section~\ref{sec:ermdp-foundation-draft} reviews entropy-regularized Markov decision processes and reinforcement learning as the single-agent basis of our formulation.
Section~\ref{sec:entropy-SMFG} introduces stationary mean field games as a large population limit of stochastic games.
Section~\ref{sec:SMFG-existence-uniqueness} establishes existence and uniqueness of equilibrium.
Section~\ref{sec:proposed-stationarity} relates nearly stationary states of our coupled dynamics to approximate equilibria.
Finally, in Sec.~\ref{sec:uniform-graph-instability}, we derive the condition for the uniform stationary state to lose stability on the graph environments considered in the main text.

\subsection{Entropy-regularized Markov decision process and reinforcement learning}
\label{sec:ermdp-foundation-draft}
We review the elements of the entropy-regularized Markov decision process (ERMDP) and reinforcement learning algorithms needed to understand the stationary mean field game and the coupled dynamics of population movement and cue modulation in this study.
Further details and biological interpretations of the formulation are provided in the SM of our previous work~\cite{Kato2025PRXLife}.

A Markov decision process (MDP) describes repeated interactions between an agent and its environment~\cite{PutermanMDP1994,BartoSutton2018}.
At each time step $t=0,1,\ldots$, the agent observes its current state $s_t\in\mathcal{S}$ and selects the next state $s_{t+1}$ according to a policy $\pi(\cdot|s_t)$, which specifies the probabilities of possible transitions.
The environment assigns a reward $r_t=r(s_t)$ that indicates how desirable the current state is to the agent.
An ERMDP additionally accounts for the cost of controlling the agent's behavior, balancing reward exploitation against control cost~\cite{Ziebart2010thesis,Haarnoja2017ICML}.
The agent seeks a policy $\pi^\dagger\in\argmax_\pi\mathcal{J}[\pi]$ that maximizes the following expected discounted cumulative reward:
\begin{subequations}
    \label{eq:ermdp-foundation-objective}
    \begin{align}
        \mathcal{J}[\pi]&\coloneqq\mathbb{E}\left[\sum_{t=0}^{\infty}\gamma^t R^\pi(s_t,s_{t+1})\right],\\
        R^\pi(s,s')&\coloneqq r(s)-\frac{1}{\beta}\log\frac{\pi(s'|s)}{p(s'|s)}.
    \end{align}
\end{subequations}
The expectation is taken over trajectories generated by $s_0\sim p_0$ and $s_{t+1}\sim\pi(\cdot|s_t)$, where $p_0\in\Delta(\mathcal{S})$ is an initial distribution.
The discount factor $0\leq\gamma<1$ gives greater weight to immediate rewards than to distant future rewards.
The second term in $R^\pi$, absent in the conventional MDP objective, penalizes deviations from an intrinsic policy $p(s'|s)$ representing the agent's behavior in the absence of control.
The inverse control weight $\beta>0$ sets the balance between reward exploitation and control cost, with smaller $\beta$ assigning a greater penalty to such deviations.

A value-based approach finds $\pi^\dagger$ indirectly by determining the maximum cumulative reward attainable from each state and using these values to construct the optimal policy.
We define the state value function $V^\pi(s)$ as the expected discounted cumulative reward obtained by following $\pi$ from state $s$, and the optimal state value $V^*(s)$ as its maximum over policies:
\begin{align}
    \label{eq:ermdp-foundation-values}
    V^\pi(s)&\coloneqq\mathbb{E}\left[\sum_{t=0}^{\infty}\gamma^t R^\pi(s_t,s_{t+1})\relmiddle|s_0=s\right],\quad s\in\mathcal{S},\\
    V^*(s)&\coloneqq\max_\pi V^\pi(s),\quad s\in\mathcal{S}.\notag
\end{align}
Thus, $\mathcal{J}[\pi]=\sum_s p_0(s)V^\pi(s)$.
The maximizer $\pi^\dagger$ of $\mathcal{J}$ can be written in terms of $V^*$:
\begin{equation}
    \label{eq:ermdp-foundation-optimal-policy}
    \pi^\dagger(s'|s)=\frac{p(s'|s)\exp(\beta\gamma V^*(s'))}{\sum_{s'\in\mathcal{S}}p(s'|s)\exp(\beta\gamma V^*(s'))},\quad s,s'\in\mathcal{S}.
\end{equation}

To find $V^*$ without evaluating every possible policy, we use the fact that it is the unique fixed point $V^*=\mathcal{B}^*[V^*]$ of the Bellman optimality operator.
This operator relates the value of the current state to the immediate reward and the values of possible next states:
\begin{equation}
    \label{eq:ermdp-foundation-optimality-operator}
    \mathcal{B}^*[V](s)\coloneqq r(s)+\frac{1}{\beta}\log\sum_{s'\in\mathcal{S}}p(s'|s)\exp(\beta\gamma V(s')),\quad s\in\mathcal{S}.
\end{equation}
The fixed-point relation is called the Bellman optimality equation and allows $V^*$ to be computed by value iteration, $V^{(n+1)}=\mathcal{B}^*[V^{(n)}]$.
Since $\mathcal{B}^*$ is a $\gamma$-contraction, $\|\mathcal{B}^*[V]-\mathcal{B}^*[W]\|_\infty\leq\gamma\|V-W\|_\infty$, this iteration converges to $V^*$.
This contraction property is also used to assess individual optimality at nearly stationary states in SM Sec.~\ref{sec:proposed-stationarity}.

In practice, however, an agent may have access to rewards only through local interactions with its environment, rather than knowing the rewards of all possible states in advance.
Reinforcement learning (RL) considers the setting in which an agent learns an optimal policy from a sequence of states and rewards $(s_t,r_t)_{t=0,1,\ldots}$ observed through interactions with its environment~\cite{BartoSutton2018}.
TD learning uses the observed reward and the current value estimate $V_t$ to approximate the unknown optimal value through the Bellman equation: $V^*(s_t)=\mathcal{B}^*[V^*](s_t)\approx\mathcal{B}^*[V_t](s_t)$.
This motivates the following update at the visited state:
\begin{subequations}
    \label{eq:ermdp-foundation-td-update}
    \begin{align}
        V_{t+1}(s_t)&=(1-\alpha_t)V_t(s_t)+\alpha_t\left(r_t+\frac{1}{\beta}\log\sum_{s'\in\mathcal{S}}p(s'|s_t)\exp(\beta\gamma V_t(s'))\right),\\
        V_{t+1}(s)&=V_t(s),\qquad s\neq s_t.
    \end{align}
\end{subequations}
where $r_t=r(s_t)$ is the observed reward and $\alpha_t\in(0,1)$ is the update rate.

Learning from observations also requires exploration of states whose rewards are not yet known.
With an initially uniform value estimate, the following greedy behavior policy supports exploration through intrinsic stochastic behavior; as learning proceeds, it favors states with higher estimated value:
\begin{equation}
    \label{eq:ermdp-foundation-learning-policy}
    \pi_t(s'|s)\coloneqq\frac{p(s'|s)\exp(\beta\gamma V_t(s'))}{\sum_{s'\in\mathcal{S}}p(s'|s)\exp(\beta\gamma V_t(s'))}.
\end{equation}
Combining this behavior policy with the value update~\eqref{eq:ermdp-foundation-td-update} gives the greedy G-learning algorithm~\cite{Fox2016}.
This algorithm inspires our coupled dynamics of population movement and cue modulation, whose derivation is given in SM Sec.~\ref{sec:algorithm-derivation}.

\subsection{From stochastic games to stationary mean field games}
\label{sec:entropy-SMFG}
We first extend the single-agent ERMDP to a stochastic game involving multiple agents.
We then introduce mean field games as the large population limit of the stochastic game with identical agents and anonymous interactions, before specializing to the stationary case.

In the ERMDP above, the desirability of each state for a single agent is determined solely by the environment.
When multiple agents interact within the same environment, however, the desirability of a state can also depend on where the other agents are.
For example, higher local population densities can facilitate cooperation or intensify competition for resources.
Stochastic games extend individual decision making in MDPs to this setting, in which each agent chooses a policy to increase its own expected cumulative reward against the other agents~\cite{Lauriere2022ArXiv}.

We first formulate a symmetric stochastic game among $N$ agents in a common state space $\mathcal{S}$, with the same reward function $r$, intrinsic policy $p$, discount factor $\gamma$, and inverse control weight $\beta$.
Their interactions are assumed to be anonymous: the reward of each agent depends on how many agents occupy each state, irrespective of their identities.
Accordingly, at time $t=0,1,\ldots$, agent $i=1,\ldots,N$ in state $s_t^i$ receives a reward $r(s_t^i;\mu_t^N)$,
where $\mu_t^N(s)\coloneqq N^{-1}\sum_{i=1}^N\mathbbm{1}_{s_t^i=s}$ represents the fraction of agents in state $s\in\mathcal{S}$.
We consider decentralized decision making: each agent chooses its next state based on its own current state at each time step.
At any given time, each agent's policy therefore has the same form as in the MDP above: $\pi:\mathcal{S}\to\Delta(\mathcal{S})$.
We denote the space of these policies by $\Pi$.
As agents move, the population density $\mu_t^N$ evolves, changing the rewards an agent can expect along its future trajectory.
Consequently, the preferred next state may depend on time even when the agent's current state is the same.
To allow for this time dependence, each agent can use a different policy at each time step, giving a policy sequence $\pi^i=(\pi_t^i)_{t\geq0}\in\Pi^{\mathbb{N}}$.
The state of agent $i$ evolves according to $s_{t+1}^i\sim\pi_t^i(\cdot|s_t^i)$.
Each agent evaluates the expected cumulative reward along its own trajectory, which also depends on the other agents' policies through the population density.
Writing $\boldsymbol{\pi}=(\pi^1,\ldots,\pi^N)$ for the policies of all agents, we define the objective of agent $i$ by
\begin{subequations}
    \label{eq:mfg-foundation-n-player-objective}
    \begin{align}
        \mathcal{J}^i[\boldsymbol{\pi}]&\coloneqq\mathbb{E}\left[\sum_{t=0}^{\infty}\gamma^t R^{\pi_t^i}(s_t^i,s_{t+1}^i;\mu_t^N)\right],\\
        R^\pi(s,s';\mu)&\coloneqq r(s;\mu)-\frac{1}{\beta}\log\frac{\pi(s'|s)}{p(s'|s)},
    \end{align}
\end{subequations}
where the expectation is taken over the trajectories of all agents, with their initial states drawn independently from a common distribution $\mu_0\in\Delta(\mathcal{S})$.
The game is characterized by the tuple $(N,\mathcal{S},p,r,\gamma,\beta,\mu_0)$.

Without a central controller, each agent seeks to maximize its own objective $\mathcal{J}^i[\boldsymbol{\pi}]$ by changing its own policy $\pi^i$.
As each agent pursues its own interests, we ask whether their policies can form an equilibrium in which no agent can improve its objective by changing its own policy while others keep their policies fixed.
Such a set of policies is called a Nash equilibrium, a solution concept in game theory.
More generally, $\boldsymbol{\pi}$ is a $\delta$-Nash equilibrium if no agent can improve its objective by more than $\delta\geq0$ through a unilateral policy change:
\begin{equation}
    \label{eq:mfg-foundation-n-player-nash}
    \mathcal{J}^i[\boldsymbol{\pi}]\geq\sup_{\pi'}\mathcal{J}^i[\pi',\pi^{-i}]-\delta,\quad i=1,\ldots,N,
\end{equation}
where $\pi^{-i}=(\pi^j)_{j\neq i}$ denotes the policies of all other agents and $\pi'=(\pi'_t)_{t\geq0}\in\Pi^{\mathbb{N}}$ is a policy sequence for agent $i$.
The case $\delta=0$ corresponds to an exact Nash equilibrium.

The effect of a unilateral policy change on the population density diminishes as \(N\) grows, because each agent contributes only a fraction \(1/N\) to that density.
This motivates the mean field approximation, which neglects the effect of an individual agent's policy change on the population density while retaining the effect of that density on the agents' rewards~\cite{Lauriere2022ArXiv}.
In the limit $N\to\infty$, a mean field game (MFG) describes an infinitesimal representative agent choosing its policy against a prescribed population mean-field $(\mu_t)_{t\geq0}$.
For a policy sequence $(\pi_t)_{t\geq0}$, this agent's objective is
\begin{equation}
    \label{eq:mfg-foundation-evolutive-objective}
    \mathcal{J}^{\mathrm{MFG}}[(\pi_t)_{t\geq0};(\mu_t)_{t\geq0}]\coloneqq\mathbb{E}\left[\sum_{t=0}^{\infty}\gamma^tR^{\pi_t}(s_t,s_{t+1};\mu_t)\right],
\end{equation}
where $s_0\sim\mu_0$ and $s_{t+1}\sim\pi_t(\cdot|s_t)$.
Although the mean-field is held fixed when an individual agent evaluates a policy change, at equilibrium it must coincide with the population distribution generated when all agents follow the same policy sequence.
Thus, a \(\delta\)-Nash equilibrium requires both individual optimality up to an error \(\delta\) and consistency between the policy sequence and the mean-field:
\begin{subequations}
    \label{eq:MFG-Nash}
    \begin{align}
        \mathcal{J}^{\mathrm{MFG}}[(\hat{\pi}_t)_{t\geq0};(\hat{\mu}_t)_{t\geq0}]&\geq\sup_{(\pi_t)_{t\geq0}}\mathcal{J}^{\mathrm{MFG}}[(\pi_t)_{t\geq0};(\hat{\mu}_t)_{t\geq0}]-\delta,\label{eq:MFG-Nash-opt}\\
        \hat{\mu}_{t+1}(s')&=\sum_{s\in\mathcal{S}}\hat{\mu}_t(s)\hat{\pi}_t(s'|s),\qquad\hat{\mu}_0=\mu_0.\label{eq:MFG-Nash-cons}
    \end{align}
\end{subequations}
The pair $((\hat{\pi}_t)_{t\geq0},(\hat{\mu}_t)_{t\geq0})$ is then called a $\delta$-Nash equilibrium of an evolutive MFG.
Under suitable assumptions, these equilibria provide approximate Nash equilibria for large finite populations, with the tolerance \(\delta\) increased by an approximation error of order $O(1/\sqrt{N})$.

In this formulation, the mean-field $\mu_t$ can change over time, and the rewards $r(s;\mu_t)$ and optimal policy $\pi_t$ can also be time-dependent even when $r$ and $p$ have no explicit time dependence.
To study the optimization and learning of time-independent policies, as in the ERMDP above, we instead evaluate an agent's policy against a population mean-field that has reached a stationary state.
The asymptotic MFG formulation evaluates a stationary policy $\pi$ for a representative agent whose initial state is drawn from \(\mu_0\), with the mean-field held constant at \(\mu_t=\mu\), \(t\geq0\).
This mean-field must coincide with the limiting population density generated when all agents follow $\pi$: $\mu=\mathcal{L}^\pi(\mu_0)\coloneqq\lim_{t\to\infty}\mu_0\pi^t$~\cite{Angiuli2022MCSS}.
If the Markov chain generated by a stationary policy \(\pi\) is ergodic, the limiting distribution is independent of the initial distribution $\mu_0$ and coincides with the unique stationary distribution \(\mu^\pi\), so that \(\mathcal{L}^\pi(\mu_0)=\mu^\pi\) and \(\mu^\pi\pi=\mu^\pi\).
Here, distributions are regarded as row vectors.
To describe a representative agent sampled from a stationary population, we instead draw its initial state from the population distribution itself, leading to the stationary MFG (SMFG) formulation~\cite{Subramanian2019AAMAS,Anahtarci2023DynGamesAppl,Angiuli2022MCSS,Guan2022ACC}.
The SMFG is characterized by the tuple $(\mathcal{S},p,r,\gamma,\beta)$.
For a fixed mean-field $\mu$ and stationary policy $\pi$, the objective is
\begin{equation}
    \label{eq:mfg-foundation-stationary-objective}
    \mathcal{J}^{\mathrm{SMFG}}[\pi;\mu]\coloneqq\mathbb{E}\left[\sum_{t=0}^{\infty}\gamma^tR^\pi(s_t,s_{t+1};\mu)\right],
\end{equation}
where the expectation is taken over trajectories generated by $s_{t+1}\sim\pi(\cdot|s_t)$ from $s_0\sim\mu$.
When evaluating a unilateral policy change, $\mu$ is held fixed both as the mean-field affecting rewards and as the initial distribution.
A stationary policy and population density $(\hat{\pi},\hat{\mu})$ constitute a $\delta$-Nash equilibrium if
\begin{subequations}
    \label{eq:SMFG-Nash}
    \begin{align}
        \mathcal{J}^{\mathrm{SMFG}}[\hat{\pi};\hat{\mu}]&\geq\sup_{\pi\in\Pi}\mathcal{J}^{\mathrm{SMFG}}[\pi;\hat{\mu}]-\delta,\label{eq:SMFG-Nash-opt}\\
        \hat{\mu}&=\mu^{\hat{\pi}}.\label{eq:SMFG-Nash-consist}
    \end{align}
\end{subequations}

For a fixed mean-field $\mu$, the objective $\mathcal{J}^{\mathrm{SMFG}}[\pi;\mu]$ coincides with the ERMDP objective~\eqref{eq:ermdp-foundation-objective} with reward $r(s;\mu)$ and initial distribution $p_0=\mu$.
The value-based approach introduced for the ERMDP therefore applies, with the state value \(V^\pi_\mu\) and optimal state value \(V^*_\mu\) now depending on the fixed mean-field \(\mu\):
\begin{align}
    \label{eq:mfg-foundation-values}
    V^\pi_\mu(s)&\coloneqq\mathbb{E}\left[\sum_{t=0}^{\infty}\gamma^tR^\pi(s_t,s_{t+1};\mu)\relmiddle|s_0=s\right],\quad s\in\mathcal{S},\\
    V^*_\mu(s)&\coloneqq\max_{\pi\in\Pi}V^\pi_\mu(s),\quad s\in\mathcal{S}.\notag
\end{align}
Averaging over the mean-field gives $\mathcal{J}^{\mathrm{SMFG}}[\pi;\mu]=\sum_s\mu(s)V^\pi_\mu(s)$.
With $\mu$ held fixed, $V^*_\mu$ is the unique solution of the associated Bellman optimality equation with reward $r(s;\mu)$.
As in the ERMDP, the Bellman operator is also a $\gamma$-contraction and value iteration converges to $V^*_\mu$.
The corresponding optimal policy is:
\begin{equation}
    \label{eq:mfg-foundation-optimal-policy}
    \pi^*_\mu(s'|s)=\frac{p(s'|s)\exp(\beta\gamma V^*_\mu(s'))}{\sum_{s'\in\mathcal{S}}p(s'|s)\exp(\beta\gamma V^*_\mu(s'))},\quad s,s'\in\mathcal{S}.
\end{equation}
An exact equilibrium can therefore be found by seeking a value--density pair $(\hat{V},\hat{\mu})$ satisfying
\begin{subequations}
    \label{eq:SMFG-Nash-via-value}
    \begin{align}
        \hat{V}&=V^*_{\hat{\mu}},\label{eq:cond-optimality}\\
        \hat{\mu}&=\hat{\mu}\pi^*_{\hat{\mu}}.
    \end{align}
\end{subequations}
For a $\delta$-Nash equilibrium, the optimality condition~\eqref{eq:SMFG-Nash-opt} can equivalently be written as $\sum_s\hat{\mu}(s)[V^*_{\hat{\mu}}(s)-V^{\hat{\pi}}_{\hat{\mu}}(s)]\leq\delta$.

Learning an SMFG equilibrium requires adapting a policy or value estimate to rewards that change as the mean-field evolves according to the policy.
Learning algorithms therefore update a policy $\pi^k$ or value estimate $V^k$, together with an estimate $\mu^k$ of the mean-field, where $k=0,1,\ldots$ are learning iterations~\cite{Lauriere2022ArXiv}.
For example, alternating methods first find a policy $\pi^{k+1}$ optimal for the current mean-field $\mu^k$, then update the mean-field to the stationary density induced when all agents follow $\pi^{k+1}$:
    $\pi^{k+1}=\pi^*_{\mu^k},\ \mu^{k+1}=\mu^{\pi^{k+1}}$.
Two-timescale methods instead gradually update $V^k$ and $\mu^k$ at every iteration, using different value and mean-field update rates $\alpha_k^V,\alpha_k^\mu\in(0,1)$~\cite{Subramanian2019AAMAS,Angiuli2022MCSS}.
For example, a TD update~\eqref{eq:ermdp-foundation-td-update} of $V^k$ with reward $r(s;\mu^k)$ and rate $\alpha_k^V$ can be coupled to a mean-field update
    $\mu^{k+1}=(1-\alpha_k^\mu)\mu^k+\alpha_k^\mu\mu^k\pi^k$,
where $\pi^k$ is the greedy policy~\eqref{eq:ermdp-foundation-learning-policy} obtained from $V^k$, and \(\mu^k\pi^k\) is the population distribution after one transition under that policy.
Two-timescale schemes targeting an MFG equilibrium typically update $\mu^{k}$ more slowly than $V^k$, with $\alpha_k^\mu/\alpha_k^V\to0$~\cite{Angiuli2022MCSS}.
This allows $V^k$ to approach the optimal value $V^*_{\mu^k}$ for the current mean-field before $\mu^k$ changes appreciably.
Our coupled dynamics are inspired by such simultaneous updates, with the endogenous cue representing a value estimate $V_t=Z_t/\beta$ and population movement updating the mean-field $\mu_t$.
The role of relative update rates in these schemes motivates our comparison of the timescales of cue modulation and population movement in the main text.
The derivation of our dynamics is given in SM Sec.~\ref{sec:algorithm-derivation}.

Density-dependent mechanical interactions such as crowding~\cite{Sadati2013Diff,Colin2019NatCommun} and adhesion~\cite{Friedl2017CSHPerspectBiol,Laganenka2016NatComm} can affect the agents' behavior even without cue guidance.
To account for such effects, we allow the intrinsic policy $p(s'|s,\mu)$ to also depend on population density, as in density-dependent random walks~\cite{Painter2002CanApplMathQuart,Anguige2009JMathBiol,Bubba2019JTheorBiol,Buttenschon2018JMathBiol}.
For fixed $\mu$, the individual problem remains an ERMDP after replacing $p(s'|s)$ by $p(s'|s,\mu)$ in the control cost.
Since the transitions allowed by \(p(\cdot|\cdot,\mu)\) can depend on \(\mu\), the admissible policies form a density-dependent set \(\Pi(\mu)\).
With this density-dependent intrinsic policy, the greedy policy for a value function $V$ and mean-field $\mu$ becomes
\begin{equation}
    \label{eq:mfg-foundation-density-dependent-policy}
    \Gamma^\pi[V,\mu](s'|s)\coloneqq\frac{p(s'|s,\mu)\exp(\beta\gamma V(s'))}{\sum_{s''\in\mathcal{S}}p(s''|s,\mu)\exp(\beta\gamma V(s''))}.
\end{equation}
Even for a fixed value estimate $V$, this policy depends on the current population density $\mu_t$, making the population update $\mu_{t+1}=\mu_t\Gamma^\pi[V,\mu_t]$ nonlinear in $\mu_t$.
For density-independent $p$, $\Gamma^\pi$ is related to the cue-dependent movement policy in Eq.~\eqref{eq:dist-G-learning-policy} by $\Gamma^\pi[V,\mu]=\pi[\beta V]$.
The nonlinear population dynamics can therefore admit several stationary densities, even if the transition matrix \(\Gamma^\pi[V,\mu]\) has a unique stationary distribution for each fixed \(\mu\).
A pair $(\hat{\pi},\hat{\mu})$ is a $\delta$-Nash equilibrium of this extended SMFG if
\begin{subequations}
    \label{eq:SMFG-Nash-general}
    \begin{align}
        \mathcal{J}^{\mathrm{SMFG}}[\hat{\pi};\hat{\mu}]&\geq\sup_{\pi\in\Pi(\hat{\mu})}\mathcal{J}^{\mathrm{SMFG}}[\pi;\hat{\mu}]-\delta,\label{eq:SMFG-Nash-opt-general}\\
        \hat{\mu}&=\hat{\mu}\hat{\pi}.\label{eq:SMFG-Nash-consist-general}
    \end{align}
\end{subequations}
For an exact equilibrium, $\hat{\pi}=\Gamma^\pi[V^*_{\hat{\mu}},\hat{\mu}]$ provides an individually optimal policy.
In principle, $\gamma$ and $\beta$ could also depend on the stationary mean-field, but we keep them fixed because we expect that comparable qualitative effects can be represented through the density dependence of $r$ and $p$.
Existence and sufficient conditions for uniqueness of an exact equilibrium are established in Propositions~\ref{prop:SMFG-general-existence} and~\ref{prop:SMFG-monotone-uniqueness}.

At a Nash equilibrium, no individual can improve its objective by changing its own policy, but a change in policy across the entire population may increase the population-level objective.
In stationary mean field control (SMFC), a central planner controls all agents and chooses their common stationary policy $\pi$ to maximize social welfare~\cite{Subramanian2019AAMAS,Angiuli2022MCSS,Lauriere2022ArXiv}:
\begin{equation}
    \label{eq:mfg-foundation-social-objective}
    \mathcal{J}^{\mathrm{SMFC}}[\pi]\coloneqq\mathbb{E}\left[\sum_{t=0}^{\infty}\gamma^tR^\pi(s_t,s_{t+1};\mu^\pi)\right]=\mathcal{J}^{\mathrm{SMFG}}[\pi;\mu^\pi],
\end{equation}
where $s_0\sim\mu^\pi$, $s_{t+1}\sim\pi(\cdot|s_t)$, and $\mu^\pi$ is the unique stationary distribution of $\pi$.
The maximizer $\pi^*\in\argmax_\pi\mathcal{J}^{\mathrm{SMFC}}[\pi]$ is called a stationary social optimum.
By definition, $\mathcal{J}^{\mathrm{SMFC}}[\pi^*]\geq\mathcal{J}^{\mathrm{SMFG}}[\hat{\pi};\hat{\mu}]$ for any stationary Nash equilibrium $(\hat{\pi},\hat{\mu})$ with $\hat{\mu}=\mu^{\hat{\pi}}$.
In SMFG, an individual optimizes $\mathcal{J}^{\mathrm{SMFG}}[\pi;\hat{\mu}]$ with $\hat{\mu}$ fixed, whereas in SMFC the planner changes the policy for the entire population, so changing $\pi$ also changes $\mu^\pi$, affecting $r(\cdot;\mu^\pi)$ and, when density-dependent, $p(\cdot|\cdot,\mu^\pi)$.
Solving SMFC thus generally requires population-level optimization, where the state is a distribution \(\mu\in\Delta(\mathcal{S})\) that takes continuous values even when the individual state space \(\mathcal{S}\) is finite.

\subsection{Existence and uniqueness of equilibrium}
\label{sec:SMFG-existence-uniqueness}
We first establish existence of a Nash equilibrium of the extended SMFG under standard continuity assumptions using Brouwer's fixed-point theorem (Prop.~\ref{prop:SMFG-general-existence}).
We then establish uniqueness for dispersal-promoting rewards and a density-independent, reversible intrinsic movement policy (Prop.~\ref{prop:SMFG-monotone-uniqueness}).

\begin{assumption}
\label{ass:SMFG-general-continuity}
The state space $\mathcal{S}$ is finite, with $n=|\mathcal{S}|$, and $0\leq\gamma<1$ and $\beta>0$.
For every $s,s'\in\mathcal{S}$, $r(s;\mu)$ and $p(s'|s,\mu)$ are continuous functions of $\mu\in\Delta(\mathcal{S})$.
\end{assumption}

Write $p_\mu$ for the transition matrix with entries $p(s'|s,\mu)$, with distributions regarded as row vectors.
Continuity of \(r\) on the compact simplex gives the finite uniform reward bound
\begin{equation}
    R\coloneqq\max_{\mu\in\Delta(\mathcal{S}),\,s\in\mathcal{S}}|r(s;\mu)|.
\end{equation}
We define the Bellman and population-update operators
\begin{align}
    \Gamma^V[V,\mu](s)&\coloneqq r(s;\mu)+\frac1\beta\log\sum_{s'\in\mathcal{S}}p(s'|s,\mu)\exp(\beta\gamma V(s')),\ s\in\mathcal{S},\\
    \Gamma^\mu[V,\mu](s)&\coloneqq\sum_{s'\in\mathcal{S}}\Gamma^\pi[V,\mu](s|s')\mu(s'),\ s\in\mathcal{S}.
\end{align}

\begin{proposition}[Existence of a Nash equilibrium of the extended SMFG]
\label{prop:SMFG-general-existence}
Under Assumption~\ref{ass:SMFG-general-continuity}, there exists a pair $(\hat{\pi},\hat{\mu})$ satisfying Eq.~\eqref{eq:SMFG-Nash-general} with $\delta=0$.
\end{proposition}

\begin{proof}
For fixed $\mu$, we apply the standard contraction argument for the Bellman operator $\Gamma^V[\cdot,\mu]$~\cite{PutermanMDP1994,Fox2016}.
Since $\Gamma^V[\cdot,\mu]$ is a $\gamma$-contraction, i.e.,
    $\|\Gamma^V[V,\mu]-\Gamma^V[W,\mu]\|_\infty
    \leq\gamma\|V-W\|_\infty$
and satisfies $\|\Gamma^V[V,\mu]\|_\infty\leq R+\gamma\|V\|_\infty$, the contraction mapping theorem gives a unique fixed point $V^*_\mu$ with $\|V^*_\mu\|_\infty\leq R/(1-\gamma)$.
Moreover, $\pi^*_{\mu}\coloneqq\Gamma^\pi[V^*_\mu,\mu]$ maximizes $\mathcal{J}^{\mathrm{SMFG}}[\pi;\mu]$ over $\pi\in\Pi(\mu)$.

To show continuity of $\mu\mapsto \pi^*_\mu$, we need to show continuity of $\mu\mapsto V^*_\mu$.
Using the fixed-point equations $V^*_\mu=\Gamma^V[V^*_\mu,\mu]$ and $V^*_\nu=\Gamma^V[V^*_\nu,\nu]$, we obtain
\begin{align*}
    \|V^*_\mu-V^*_\nu\|_\infty
    &=\|\Gamma^V[V^*_\mu,\mu]-\Gamma^V[V^*_\nu,\nu]\|_\infty
    \nonumber\\
    &\leq\|\Gamma^V[V^*_\mu,\mu]-\Gamma^V[V^*_\nu,\mu]\|_\infty
    +\|\Gamma^V[V^*_\nu,\mu]-\Gamma^V[V^*_\nu,\nu]\|_\infty
    \nonumber\\
    &\leq\gamma\|V^*_\mu-V^*_\nu\|_\infty
    +\|\Gamma^V[V^*_\nu,\mu]-\Gamma^V[V^*_\nu,\nu]\|_\infty.
\end{align*}
Rearranging gives
\begin{equation*}
    \|V^*_\mu-V^*_\nu\|_\infty
    \leq\frac{\|\Gamma^V[V^*_\nu,\mu]-\Gamma^V[V^*_\nu,\nu]\|_\infty}{1-\gamma}.
\end{equation*}
For fixed $\nu$, continuity of $r$ and $p$ makes the right-hand side tend to zero as $\mu\to\nu$.
Thus $\mu\mapsto V^*_\mu$ is continuous.

Continuity of $p_\mu$ and $V^*_\mu$ gives continuity of $\mu\mapsto \pi^*_\mu$.
Thus
\begin{equation}
    \label{eq:SMFG-general-fixed-point-map}
    F(\mu)\coloneqq\mu \pi^*_\mu=\Gamma^\mu[V^*_\mu,\mu]
\end{equation}
is a continuous map from the compact convex simplex into itself.
By Brouwer's fixed point theorem, some $\hat{\mu}$ satisfies $\hat{\mu}=F(\hat{\mu})$.
Setting $\hat{\pi}=\pi^*_{\hat{\mu}}$ satisfies both equilibrium conditions in Eqs.~\eqref{eq:SMFG-Nash-general} with $\delta=0$.
\end{proof}

Although uniqueness can also be guaranteed by showing that the map $F$ in Eq.~\eqref{eq:SMFG-general-fixed-point-map} is a contraction,
this approach typically requires sufficiently weak dependence of the reward on the mean-field.
Instead, for a density-independent, reversible intrinsic movement policy, as used in our numerical experiments, we establish uniqueness under the Lasry--Lions monotonicity condition~\cite{Lasry-Lions2007JJM,Lauriere2022ArXiv}.
In particular, this condition holds for dispersal-promoting rewards of the form $r(s;\mu)=g(s)-f_s(\mu(s))$, where each $f_s$ is continuous and nondecreasing.

\begin{proposition}[Uniqueness under reward monotonicity]
\label{prop:SMFG-monotone-uniqueness}
Suppose Assumption~\ref{ass:SMFG-general-continuity} holds and $p(s'|s,\mu)=p(s'|s)$ is irreducible and reversible with respect to a probability distribution $\rho$ with $\rho(s)>0$, that is,
\begin{equation}
    \rho(s)p(s'|s)=\rho(s')p(s|s')
    \qquad\text{for all }s,s'\in\mathcal{S}.
\end{equation}
If the reward function $r(s;\mu)$ satisfies the Lasry--Lions monotonicity condition
\begin{equation}
    \label{eq:SMFG-reward-monotonicity}
    \sum_{s\in\mathcal{S}}[\mu(s)-\nu(s)][r(s;\mu)-r(s;\nu)]\leq0
    \qquad\text{for all }\mu,\nu\in\Delta(\mathcal{S}),
\end{equation}
then Eq.~\eqref{eq:SMFG-Nash-general} has exactly one equilibrium pair $(\hat{\pi},\hat{\mu})$ for $\delta=0$ with strictly positive $\hat{\mu}$.
\end{proposition}

\begin{proof}
For any exact equilibrium $(\pi,\mu)$, the Bellman equations, together with the optimality and consistency conditions in Eq.~\eqref{eq:SMFG-Nash-general}, give
\begin{equation*}
    \label{eq:SMFG-general-stationary-gap}
    (1-\gamma)\sum_s\mu(s)[V^*_\mu(s)-V^\pi_\mu(s)]
    =\frac{1}{\beta}\sum_s\mu(s)
    D_{\mathrm{KL}}\!\left(\pi(\cdot|s)\|\pi^*_\mu(\cdot|s)\right)=0.
\end{equation*}
Hence $\pi=\pi^*_\mu$ on the support of $\mu$, which also implies $\mu\pi^*_\mu=\mu$.
Since $\pi^*_\mu$ has the same support as the irreducible kernel $p$, $\mu$ is strictly positive and $\pi=\pi^*_\mu$.
If $\gamma=0$, $\pi^*_\mu=p$ for every $\mu$, so uniqueness follows immediately from irreducibility.
Henceforth suppose $0<\gamma<1$.

Consider two equilibria $(\pi_1,\mu_1)$ and $(\pi_2,\mu_2)$.
The Bellman equation for $V^*_{\mu_i}$ gives
\begin{equation*}
    \log\sum_jp(j|s)\exp(\beta\gamma V^*_{\mu_i}(j))
    =\beta[V^*_{\mu_i}(s)-r(s;\mu_i)].
\end{equation*}
Substituting this expression for the normalization term in the optimal movement policy yields, whenever $p(s'|s)>0$,
\begin{align*}
    \log\pi_i(s'|s)
    &=\log p(s'|s)+\beta\gamma V^*_{\mu_i}(s')
    -\log\sum_jp(j|s)\exp(\beta\gamma V^*_{\mu_i}(j))\\
    &=\log p(s'|s)+\beta\gamma V^*_{\mu_i}(s')
    -\beta[V^*_{\mu_i}(s)-r(s;\mu_i)].
\end{align*}
Subtracting the expressions for $i=1$ and $i=2$ gives
\begin{equation*}
    \log\frac{\pi_1(s'|s)}{\pi_2(s'|s)}
    =\beta\gamma[V^*_{\mu_1}(s')-V^*_{\mu_2}(s')]
    -\beta[V^*_{\mu_1}(s)-V^*_{\mu_2}(s)]
    +\beta[r(s;\mu_1)-r(s;\mu_2)].
\end{equation*}
Averaging this identity with respect to $\mu_1(s)\pi_1(s'|s)$ and using stationarity $\sum_s\mu_1(s)\pi_1(s'|s)=\mu_1(s')$, we obtain
\begin{align*}
    \frac1\beta\sum_s\mu_1(s)D_{\mathrm{KL}}\!\left(\pi_1(\cdot|s)\|\pi_2(\cdot|s)\right)
    &=\gamma\sum_{s,s'}\mu_1(s)\pi_1(s'|s)[V^*_{\mu_1}(s')-V^*_{\mu_2}(s')]
    -\langle\mu_1,V^*_{\mu_1}-V^*_{\mu_2}\rangle
    +\langle\mu_1,r(\cdot;\mu_1)-r(\cdot;\mu_2)\rangle\\
    &=-(1-\gamma)\langle\mu_1,V^*_{\mu_1}-V^*_{\mu_2}\rangle
    +\langle\mu_1,r(\cdot;\mu_1)-r(\cdot;\mu_2)\rangle,
\end{align*}
where $\langle a,b\rangle=\sum_s a(s)b(s)$ denotes the inner product.
Similarly, exchanging $(\pi_1,\mu_1)$ and $(\pi_2,\mu_2)$ gives
\begin{equation*}
    \frac1\beta\sum_s\mu_2(s)D_{\mathrm{KL}}\!\left(\pi_2(\cdot|s)\|\pi_1(\cdot|s)\right)
    =(1-\gamma)\langle\mu_2,V^*_{\mu_1}-V^*_{\mu_2}\rangle
    -\langle\mu_2,r(\cdot;\mu_1)-r(\cdot;\mu_2)\rangle.
\end{equation*}
Adding these two equalities and rearranging yields
\begin{align}
    \label{eq:SMFG-monotone-equilibrium-comparison}
    \langle\mu_1-\mu_2,r(\cdot;\mu_1)-r(\cdot;\mu_2)\rangle
    &=(1-\gamma)\langle\mu_1-\mu_2,V^*_{\mu_1}-V^*_{\mu_2}\rangle\nonumber\\
    &\quad+\frac1\beta\sum_s\left[
    \mu_1(s)D_{\mathrm{KL}}\!\left(\pi_1(\cdot|s)\|\pi_2(\cdot|s)\right)
    +\mu_2(s)D_{\mathrm{KL}}\!\left(\pi_2(\cdot|s)\|\pi_1(\cdot|s)\right)\right].
\end{align}
The left-hand side is nonpositive by assumption~\eqref{eq:SMFG-reward-monotonicity}, and the KL terms on the right-hand side are nonnegative.
It remains to show that $\langle\mu_1-\mu_2,V^*_{\mu_1}-V^*_{\mu_2}\rangle\geq0$.

For any $V$, write $\pi_V(s'|s)=p(s'|s)\exp(\beta\gamma V(s'))/\sum_jp(j|s)\exp(\beta\gamma V(j))$.
Reversibility of $p$ gives $\mu_V(s)\pi_V(s'|s)=\mu_V(s')\pi_V(s|s')$ for the unique invariant distribution of $\pi_V$:
\begin{equation*}
    \mu_V(s)=\frac{\rho(s)\exp(\beta\gamma V(s))\sum_jp(j|s)\exp(\beta\gamma V(j))}
    {\sum_{i,j}\rho(i)p(j|i)\exp(\beta\gamma[V(i)+V(j)])}.
\end{equation*}
Moreover, $\mu_V=\nabla_V\Phi(V)$, where
\begin{equation*}
    \Phi(V)\coloneqq\frac{1}{2\beta\gamma}
    \log\sum_{i,j}\rho(i)p(j|i)\exp(\beta\gamma[V(i)+V(j)]).
\end{equation*}
Since $\Phi$ is a convex function, its gradient is monotone:
\begin{equation}
    \label{eq:SMFG-reversible-density-monotonicity}
    \langle\mu_V-\mu_W,V-W\rangle\geq0.
\end{equation}
Since $\pi_i=\pi^*_{\mu_i}$, we have $\mu_i=\mu_{V^*_{\mu_i}}$ for $i=1,2$, so Eq.~\eqref{eq:SMFG-reversible-density-monotonicity} gives $\langle\mu_1-\mu_2,V^*_{\mu_1}-V^*_{\mu_2}\rangle\geq0$.
Hence every term on the right-hand side of Eq.~\eqref{eq:SMFG-monotone-equilibrium-comparison} is nonnegative.
Therefore, the KL divergence term in Eq.~\eqref{eq:SMFG-monotone-equilibrium-comparison} vanishes, and positivity of $\mu_i$ gives $\pi_1=\pi_2$.
Irreducibility of this common movement policy then gives $\mu_1=\mu_2$, proving uniqueness.
\end{proof}


\subsection{Stationarity of our coupled dynamics and approximate Nash equilibria}
\label{sec:proposed-stationarity}
We relate nearly stationary states $(Z_t,\mu_t)$ under our coupled dynamics to individual optimality and population consistency defining Nash equilibria~\eqref{eq:SMFG-Nash-general}.
Writing $\pi_t=\Gamma^\pi[Z_t/\beta,\mu_t]$, Eqs.~\eqref{eq:mean-field-evolution} and~\eqref{eq:dist-G-learnSMFG-diff} become
\begin{subequations}
    \label{eq:proposed-value-population-update}
    \begin{align}
        Z_{t+1}(s)-Z_t(s)
        &=\alpha\mu_t(s)\bigl(\beta\Gamma^V[Z_t/\beta,\mu_t](s)-Z_t(s)\bigr)-D(LZ_t)(s),\\
        \mu_{t+1}&=\Gamma^\mu[Z_t/\beta,\mu_t]=\mu_t\pi_t.
    \end{align}
\end{subequations}
Recall that $0\leq\gamma<1$, $\beta>0$, $\alpha>0$, and $D\geq0$, and that the state space is finite and the reward is bounded.
For the graph Laplacian $L$, write $d_{\max}\coloneqq\max_s L_{ss}$, so $\|LZ\|_\infty\leq2d_{\max}\|Z\|_\infty$.

Proposition~\ref{prop:proposed-near-stationarity} presents a bound on the possible improvement in the objective in the optimality condition~\eqref{eq:SMFG-Nash-opt-general}.
This bound implies that if the dynamics are nearly stationary, then the states approximately satisfy the equilibrium condition~\eqref{eq:SMFG-Nash-general} as long as diffusion $D$ is sufficiently small and the population density $\mu$ is sufficiently spread over the state space.
These restrictions arise from cue diffusion and density-weighted cue modulation, both introduced to give the dynamics a physical interpretation in the derivation of Eq.~\eqref{eq:proposed-value-population-update}.
Biologically, a regime of small cue diffusion can be interpreted as haptotactic movement guided by self-deposited extracellular matrix (ECM), which remains bound to the substrate and does not diffuse~\cite{Sengupta2021NatRevMolCellBiol}. 
Furthermore, Corollary~\ref{cor:proposed-nearby-Nash} shows that a $\delta$-Nash equilibrium exists in the neighborhood of nearly stationary states under additional regularity and mixing conditions.

\begin{proposition}[Optimality bounds at nearly stationary states]
\label{prop:proposed-near-stationarity}
At a time $t$, suppose that, for some $m>0$ and $B<\infty$,
\begin{equation*}
    \min_s\mu_t(s)\geq m,\qquad \|Z_t\|_\infty\leq B,\qquad
    \|Z_{t+1}-Z_t\|_\infty\leq\epsilon^Z.
\end{equation*}
Define
\begin{equation*}
    \label{eq:proposed-Bellman-residual-bound}
    \delta_t\coloneqq\frac{2(\epsilon^Z+D\|LZ_t\|_\infty)}{\beta(1-\gamma)\alpha m}\le \frac{2(\epsilon^Z+2Dd_{\max}B)}{\beta(1-\gamma)\alpha m}.
\end{equation*}
Then, $\|Z_t-\beta V^*_{\mu_t}\|_\infty\leq \beta\delta_t/2$ and
\begin{equation}
    \label{eq:proposed-near-stationarity-optimality-bound}
    \sup_{\pi\in\Pi(\mu_t)}\mathcal{J}^{\mathrm{SMFG}}[\pi;\mu_t]
    -\mathcal{J}^{\mathrm{SMFG}}[\pi_t;\mu_t]\leq\delta_t.
\end{equation}
In particular, if $\|\mu_{t+1}-\mu_t\|_1=0$, $(\pi_t,\mu_t)$ is a $\delta_t$-Nash equilibrium of Eq.~\eqref{eq:SMFG-Nash-general}.
\end{proposition}

\begin{proof}
    We prove Eq.~\eqref{eq:proposed-near-stationarity-optimality-bound} by bounding $\|V^*_{\mu_t}-V^{\pi_t}_{\mu_t}\|_\infty\leq \|Z_t/\beta-V^*_{\mu_t}\|_\infty+\|Z_t/\beta-V^{\pi_t}_{\mu_t}\|_\infty$.
\begin{equation*}
    \Gamma^V[Z_t/\beta,\mu_t](s)-Z_t(s)/\beta
    =\frac{Z_{t+1}(s)-Z_t(s)+D(LZ_t)(s)}{\beta\alpha\mu_t(s)}\qquad s\in\mathcal{S}.
\end{equation*}
Hence $\|\Gamma^V[Z_t/\beta,\mu_t]-Z_t/\beta\|_\infty\leq b_t$ with $b_t=(\epsilon^Z+D\|LZ_t\|_\infty)/(\beta\alpha m)$.
Since $\Gamma^V[\cdot,\mu_t]$ is a $\gamma$-contraction with a fixed point $V^*_{\mu_t}$, we have
    \(\|Z_t/\beta-V^*_{\mu_t}\|_\infty
    \leq b_t+\gamma\|Z_t/\beta-V^*_{\mu_t}\|_\infty\),
hence $\|Z_t-\beta V^*_{\mu_t}\|_\infty\leq\beta b_t/(1-\gamma)$.

For a fixed policy $\pi$ and mean field $\mu$, define its policy-evaluation operator by
\begin{equation*}
    \mathcal{B}^\pi_\mu[V](s)
    \coloneqq r_g(s;\mu)
    -\frac{1}{\beta}D_{\mathrm{KL}}\!\left(\pi(\cdot|s)\|p(\cdot|s,\mu)\right)
    +\gamma\sum_{s'}\pi(s'|s)V(s') \qquad s\in\mathcal{S}.
\end{equation*}
This operator is a $\gamma$-contraction with fixed point $V^\pi_\mu$.
Since $\pi_t=\Gamma^\pi[Z_t/\beta,\mu_t]$ maximizes $\mathcal{B}^\pi_{\mu_t}[Z_t/\beta]$ state by state, we obtain
$\mathcal{B}^{\pi_t}_{\mu_t}[Z_t/\beta]=\Gamma^V[Z_t/\beta,\mu_t]$.
The same argument therefore gives $\|Z_t-\beta V^{\pi_t}_{\mu_t}\|_\infty\leq\beta b_t/(1-\gamma)$, and consequently
\begin{equation*}
    0\leq V^*_{\mu_t}(s)-V^{\pi_t}_{\mu_t}(s)
    \leq\frac{2b_t}{1-\gamma}
    \qquad s\in\mathcal{S}.
\end{equation*}
Averaging over the initial distribution $\mu_t$ proves the bound~\eqref{eq:proposed-near-stationarity-optimality-bound}.
When $\|\mu_{t+1}-\mu_t\|_1=\|\mu_t\pi_t-\mu_t\|_1=0$, $\mu_t\pi_t=\mu_t$, so the consistency condition holds.
\end{proof}

Under the following regularity assumptions and a mixing condition, a \(\delta\)-Nash equilibrium exists near a nearly stationary state.
The mixing condition requires contraction after a finite number of population updates with the cue field held fixed, extending a mixing property of finite irreducible and aperiodic Markov chains to density-dependent movement.

\begin{assumption}
\label{ass:SMFG-general-lipschitz}
$r_g$ and $p$ are Lipschitz continuous with Lipschitz constants $\ell_r,\ell_p\geq0$ such that, for all $\mu,\nu\in\Delta(\mathcal{S})$,
\begin{align}
    \|r_g(\cdot;\mu)-r_g(\cdot;\nu)\|_\infty
    &\leq\ell_r\|\mu-\nu\|_1,
    \label{eq:SMFG-r-Lipschitz}\\
    \max_s\sum_{s'}|p(s'|s,\mu)-p(s'|s,\nu)|
    &\leq\ell_p\|\mu-\nu\|_1.
\end{align}
In addition, for some integer $k\geq1$ and for every finite value field $V$, the map $F_V(\mu)\coloneqq\Gamma^\mu[V,\mu]$ satisfies
\begin{equation}
    \|F_V^{\circ j}(\mu)-F_V^{\circ j}(\nu)\|_1
    \leq\kappa_{V,j}\|\mu-\nu\|_1
    \quad\text{for all }\mu,\nu\in\Delta(\mathcal S),\quad 0\leq j\leq k,
\end{equation}
where $\kappa_{V,0}=1$, the constants $\kappa_{V,j}$ are finite and nonnegative, and $\kappa_{V,k}<1$.
\end{assumption}

\begin{corollary}[A nearby $\delta$-Nash equilibrium]
\label{cor:proposed-nearby-Nash}
Suppose the conditions of Proposition~\ref{prop:proposed-near-stationarity} and Assumption~\ref{ass:SMFG-general-lipschitz} hold, and that $\|\mu_{t+1}-\mu_t\|_1\leq\epsilon^\mu$.
Write $\kappa_{t,j}\coloneqq\kappa_{Z_t/\beta,j}$ for $0\leq j\leq k$.
Define
\begin{equation*}
    h_t\coloneqq\min\left\{2,\,
    \frac{\epsilon^\mu\sum_{j=0}^{k-1}\kappa_{t,j}}{1-\kappa_{t,k}}\right\},\qquad
    C_B\coloneqq\ell_r+\frac{\exp(2\gamma B)\ell_p}{\beta}.
\end{equation*}
Then the map $F_{Z_t/\beta}$ has a unique fixed point $\bar{\mu}_t$ satisfying
\begin{equation}
    \label{eq:proposed-density-correction}
    \|\bar{\mu}_t-\mu_t\|_1\leq h_t.
\end{equation}
The pair $(\bar{\pi}_t,\bar{\mu}_t)$, where $\bar{\pi}_t=\Gamma^\pi[Z_t/\beta,\bar{\mu}_t]$, is a $\bar{\delta}_t$-Nash equilibrium with
\begin{equation}
    \label{eq:proposed-Nash-error}
    \bar{\delta}_t\coloneqq \delta_t+\frac{2C_Bh_t}{1-\gamma}.
\end{equation}
\end{corollary}

\begin{proof}
The contraction mapping theorem gives a unique fixed point $\bar{\mu}_t$ of $F_{Z_t/\beta}^{\circ k}$.
Since
\begin{equation*}
    F_{Z_t/\beta}^{\circ k}(F_{Z_t/\beta}(\bar{\mu}_t))
    =F_{Z_t/\beta}(F_{Z_t/\beta}^{\circ k}(\bar{\mu}_t))
    =F_{Z_t/\beta}(\bar{\mu}_t),
\end{equation*}
uniqueness implies $F_{Z_t/\beta}(\bar{\mu}_t)=\bar{\mu}_t$.
Conversely, every fixed point of $F_{Z_t/\beta}$ is fixed by $F_{Z_t/\beta}^{\circ k}$, so $\bar{\mu}_t$ is also the unique fixed point of $F_{Z_t/\beta}$.
Since $F_{Z_t/\beta}^{\circ k}(\bar{\mu}_t)=\bar{\mu}_t$, the triangle inequality and the bounds on the iterates give
\begin{align*}
    \|\mu_t-\bar{\mu}_t\|_1
    &\leq\|\mu_t-F_{Z_t/\beta}^{\circ k}(\mu_t)\|_1
    +\|F_{Z_t/\beta}^{\circ k}(\mu_t)-F_{Z_t/\beta}^{\circ k}(\bar{\mu}_t)\|_1\\
    &\leq\sum_{j=0}^{k-1}
    \|F_{Z_t/\beta}^{\circ j}(F_{Z_t/\beta}(\mu_t))-F_{Z_t/\beta}^{\circ j}(\mu_t)\|_1
    +\kappa_{t,k}\|\mu_t-\bar{\mu}_t\|_1\\
    &\leq\epsilon^\mu\sum_{j=0}^{k-1}\kappa_{t,j}
    +\kappa_{t,k}\|\mu_t-\bar{\mu}_t\|_1.
\end{align*}
Rearranging and using $\|\mu_t-\bar{\mu}_t\|_1\le 2$ gives Eq.~\eqref{eq:proposed-density-correction}.

To bound the optimality error at the stationary density $\bar{\mu}_t$, we bound $\|\Gamma^V[Z_t/\beta,\mu_t]-\Gamma^V[Z_t/\beta,\bar{\mu}_t]\|_\infty$. 
Since $\|Z_t\|_\infty\leq B$, we have
\(    \exp(-\gamma B)\leq\sum_{s'}p(s'|s,\mu)\exp(\gamma Z_t(s'))\leq\exp(\gamma B)\).
Using $|\log a-\log b|\leq|a-b|/\min\{a,b\}$ for $a,b>0$, we obtain
\begin{align*}
    \left|\log\frac{\sum_{s'}p(s'|s,\mu)\exp(\gamma Z_t(s'))}{\sum_{s'}p(s'|s,\nu)\exp(\gamma Z_t(s'))}\right|
    &\leq\exp(\gamma B)\left|\sum_{s'}\bigl(p(s'|s,\mu)-p(s'|s,\nu)\bigr)\exp(\gamma Z_t(s'))\right|\\
    &\leq\exp(2\gamma B)\sum_{s'}|p(s'|s,\mu)-p(s'|s,\nu)|\\
    &\leq\exp(2\gamma B)\ell_p\|\mu-\nu\|_1.
\end{align*}
Combining this with the reward Lipschitz bound~\eqref{eq:SMFG-r-Lipschitz} and taking $\mu=\mu_t$ and $\nu=\bar{\mu}_t$ gives
\begin{equation*}
    \|\Gamma^V[Z_t/\beta,\mu_t]-\Gamma^V[Z_t/\beta,\bar{\mu}_t]\|_\infty
    \leq C_B\|\mu_t-\bar{\mu}_t\|_1.
\end{equation*}
Thus
\begin{equation*}
    \|\Gamma^V[Z_t/\beta,\bar{\mu}_t]-Z_t/\beta\|_\infty
    \leq b_t+C_Bh_t,
\end{equation*}
where $b_t$ is defined in the proof of Proposition~\ref{prop:proposed-near-stationarity}.
Applying the same argument as in Proposition~\ref{prop:proposed-near-stationarity} at $\bar{\mu}_t$ gives the bound~\eqref{eq:proposed-near-stationarity-optimality-bound} with $\delta_t$ replaced by $\bar{\delta}_t$.
By construction, $\bar{\mu}_t\bar{\pi}_t=\bar{\mu}_t$, so population consistency holds exactly.
\end{proof}

For the population update in the numerical experiments, $F_{Z_t/\beta}^{\circ j}(\mu)=\mu\pi_t^j$, so we may take $\kappa_{t,j}\le 1$.
Since every vertex of the graph can reach the root $s_0$ in at most four steps and $\pi_t(s|s)>0$ for all $s$, $\pi_t^4(s_0|s)>0$ for all $s$.
This gives the mixing bound~\cite[Sec.~4.4]{Levin2017}:
\begin{equation*}
    \kappa_{t,4}\coloneqq\frac12\max_{s,u}
    \sum_{s'}|\pi_t^4(s'|s)-\pi_t^4(s'|u)|
    =1-\min_{s,u}\sum_{s'}\min\{\pi_t^4(s'|s),\pi_t^4(s'|u)\}
    \leq1-\min_s\pi_t^4(s_0|s)<1.
\end{equation*}
Corollary~\ref{cor:proposed-nearby-Nash} consequently applies to the nearly stationary states in the numerical experiments with $k=4$.

\subsection{Instability threshold of the uniform stationary state on a finite graph}
\label{sec:uniform-graph-instability}
The jumps in the final-pattern statistics in Figs.~\ref{fig:fast-modulation}(g-j) and \ref{fig:slow-fast}(f-i) indicate a transition from uniform to non-uniform patterns.
The observed transition occurs near the loss of stability of the uniform stationary state.
We analytically derive its linear stability condition for the finite graph used in the numerical experiments when reward is local and spatially homogenous $r_g(s;\mu)=Af(\mu(s))$.

Let the graph have $n=|\mathcal{S}|$ vertices and be connected and undirected, with symmetric Laplacian $L$ and intrinsic movement matrix $p=I-\varepsilon L$.
Recall that $p$ is stochastic, $\varepsilon>0$, $\beta>0$, $0<\gamma<1$, $\alpha>0$, and $D>0$.
The coupled dynamics~\eqref{eq:mean-field-evolution} and \eqref{eq:dist-G-learnSMFG-diff} are written in vector form as
\begin{align}
 Z^+&=Z+\alpha\operatorname{diag}(\mu)
 \left[\beta Af(\mu)+\log(p\exp({\gamma Z}))-Z\right]-DLZ,\nonumber\\
 \mu^+&=\pi[Z]^\top\mu,\qquad
 \pi[Z]_{ij}=\frac{p_{ij}\exp({\gamma Z_j})}{(p\exp({\gamma Z}))_i},
 \label{eq:uniform-graph-map}
\end{align}
where exponential, logarithm, and reward act componentwise.
The uniform stationary pattern of Eq.~\eqref{eq:uniform-graph-map} is
\begin{equation}
 \mu_*=\frac1n\mathbf1,\qquad
 Z_*=\frac{\beta Af(1/n)}{1-\gamma}\mathbf1.
 \label{eq:uniform-graph-stationary-pair}
\end{equation}

We consider linear stability of the uniform stationary state~\eqref{eq:uniform-graph-stationary-pair} on the subspace of mass-conserving perturbations
\begin{equation}
 \mathcal{T}\coloneqq
 \left\{(\delta Z,\delta\mu)\in\mathbb{R}^n\times\mathbb{R}^n:
 \mathbf1^\top\delta\mu=0\right\}.
 \label{eq:uniform-graph-perturbation-space}
\end{equation}
We determine the critical threshold of $A$ at which the uniform stationary state becomes unstable under this perturbation $\mathcal{T}$.

\begin{proposition}[Uniform-state instability threshold on the graph]
\label{prop:uniform-graph-critical-A}
Denote $0=\lambda_1<\lambda_2\le\cdots\le\lambda_n$ for the Laplacian eigenvalues of the graph.
Suppose $f'(1/n)>0$, $A\ge0$, for every $j\ge2$,
$0\le1-\varepsilon\lambda_j<1,
 0\le 1-\frac{\alpha}{n}\left[1-\gamma(1-\varepsilon\lambda_j)\right]
 -D\lambda_j$, and $1-\alpha(1-\gamma)/n<1$.
Define
\begin{equation}
 A_{\mathrm c}(\lambda)=
 \frac{[1-\gamma+\gamma\varepsilon\lambda]/n+(D/\alpha)\lambda}
 {\beta\gamma f'(1/n)(2-\varepsilon\lambda)/n^2}.
 \label{eq:uniform-graph-modal-critical-A}
\end{equation}
The uniform stationary state is linearly asymptotically stable on $\mathcal{T}$ if $A<A_{\mathrm c}(\lambda_2)$ and linearly unstable on $\mathcal{T}$ if $A>A_{\mathrm c}(\lambda_2)$.
\end{proposition}

The threshold depends on graph connectivity through $\lambda_2$ and on cue diffusion relative to modulation through $D/\alpha$.
In particular, graphs with well-separated clusters have small $\lambda_2$, which lowers the threshold for instability.
The threshold is identical for fast and slow cue modulation with the same $D/\alpha$ ratio, as in our numerical setting.

For the numerical settings, $f(x)=\frac{(x/K)^2}{1+(x/K)^{10}}$, $K=0.072$, $\beta=1$, $\gamma=0.8$, $\varepsilon=0.05$, and $D/\alpha=0.025$, with $(\alpha,D)=(0.5,0.0125)$ and $(0.005,0.000125)$ for fast and slow cue modulation, respectively.
The condition of Proposition~\ref{prop:uniform-graph-critical-A} is satisfied for the graph used in the numerical experiments, with $\lambda_2\approx0.0403$, calculated below, yielding $A_{\mathrm c}(\lambda_2)\approx0.8509$.
This threshold is consistent with the numerical results in Figs.~\ref{fig:fast-modulation}(g-j) and \ref{fig:slow-fast}(f-i), where the uniform stationary state is stable for $A=0.8$ and unstable for $A=0.9$.

\begin{proof}
    [\bfseries Proof of Proposition~\ref{prop:uniform-graph-critical-A}]
Consider small perturbations $(\delta Z,\delta\mu)\in\mathcal{T}$, where $\delta Z=Z-Z_*$ and $\delta\mu=\mu-\mu_*$, retaining only first-order terms below.
At the uniform stationary state $(Z_*,\mu_*)$, $\pi[Z_*]=p$ and $\beta Af(\mu_*)+\log(pe^{\gamma Z_*})-Z_*=0$.
The first-order variation of the cue update is $(\gamma p-I)\delta Z+\beta Af'(1/n)\delta\mu$.
For the population update, differentiating the normalized movement policy gives
\begin{equation*}
 \delta\pi_{ij}=\gamma p_{ij}
 \left[\delta Z_j-(p\delta Z)_i\right],\quad i,j\in\mathcal{S}.
 \label{eq:uniform-graph-policy-variation}
\end{equation*}
Using the symmetry and stochasticity of $p$ gives
\begin{align*}
 [(\delta\pi)^\top\mu_*]_j
 &=\frac{\gamma}{n}\sum_i p_{ij}
 \left[\delta Z_j-(p\delta Z)_i\right]\\
 &=\frac{\gamma}{n}[(I-p^2)\delta Z]_j.
\end{align*}
Combining this term with $p^\top\delta\mu=p\delta\mu$ yields the full linearized update
\begin{equation}
 \begin{pmatrix}\delta Z^+\\\delta\mu^+\end{pmatrix}
 =\begin{pmatrix}
 I-\frac{\alpha}{n}(I-\gamma p)-DL&\frac{\alpha\beta A}{n}f'(1/n)I\\
 \frac{\gamma}{n}(I-p^2)&p
 \end{pmatrix}
 \begin{pmatrix}\delta Z\\\delta\mu\end{pmatrix}.
 \label{eq:uniform-graph-linearization}
\end{equation}

Choose an orthonormal Laplacian eigenbasis $u_j$, with $Lu_j=\lambda_j u_j$ and $u_1=\mathbf1/\sqrt n$.
Since $p=I-\varepsilon L$, these vectors also satisfy $pu_j=(1-\varepsilon\lambda_j)u_j$.
Expand $\delta Z=\sum_j\delta Z_j u_j$ and $\delta\mu=\sum_j\delta\mu_j u_j$, where $\delta Z_j=u_j^\top\delta Z$ and $\delta\mu_j=u_j^\top\delta\mu$ denote modal coefficients.
The cue and density coefficients within each mode therefore evolve as
\begin{align}
 \begin{pmatrix}\delta Z_j^+\\\delta\mu_j^+\end{pmatrix}
 &=M_j\begin{pmatrix}\delta Z_j\\\delta\mu_j\end{pmatrix},
 \quad M_j=\begin{pmatrix}
 a_j&\frac{\alpha\beta A}{n}f'(1/n)\\
 \frac{\gamma}{n}[1-(1-\varepsilon\lambda_j)^2]&1-\varepsilon\lambda_j
 \end{pmatrix},
 \label{eq:uniform-graph-modal-matrix}
\end{align}
where $a_j=1-\frac{\alpha}{n}[1-\gamma(1-\varepsilon\lambda_j)]-D\lambda_j$.

For the constant spatial mode $j=1$, $\lambda_1=0$ and
\begin{equation*}
 M_1=\begin{pmatrix}
 1-\frac{\alpha}{n}(1-\gamma)&\frac{\alpha\beta A}{n}f'(1/n)\\
 0&1
 \end{pmatrix}.
 \label{eq:uniform-graph-constant-mode}
\end{equation*}
For perturbations in $\mathcal{T}$, the constant density coefficient satisfies $\delta\mu_1=u_1^\top\delta\mu=\mathbf1^\top\delta\mu/\sqrt n=0$.
The cue coefficient of the constant-mode is also stable since $1-\alpha(1-\gamma)/n<1$.

For each $j\ge2$, the eigenvalues of the discrete linearized update $M_j$ are
\begin{equation}
 z_{j,\pm}=\frac12\Bigl\{a_j+1-\varepsilon\lambda_j
 \pm\sqrt{(a_j-1+\varepsilon\lambda_j)^2
 +\frac{4\alpha\beta\gamma A}{n^2}f'(1/n)
 [1-(1-\varepsilon\lambda_j)^2]}\Bigr\}.
 \label{eq:uniform-graph-modal-growth}
\end{equation}
The assumptions imply that $M_j$ is nonnegative and both diagonal entries are less than one, so its largest absolute eigenvalue is $z_{j,+}$.
$z_{j,+}>1$ holds if and only if
\begin{equation*}
 (1-a_j)\varepsilon\lambda_j
 <\frac{\alpha\beta\gamma A}{n^2}f'(1/n)
 [1-(1-\varepsilon\lambda_j)^2].
\end{equation*}
Substituting $a_j$ and cancelling $\alpha\varepsilon\lambda_j>0$ gives $A>A_{\mathrm c}(\lambda_j)$, with the reverse strict inequality giving $z_{j,+}<1$.
$A_{\mathrm c}(\lambda)$ is strictly increasing in $\lambda$, and the first unstable spatial mode has eigenvalue $\lambda_2$.
\end{proof}

To evaluate the instability threshold for a given graph, we need its algebraic connectivity $\lambda_2$.
For the tree graph used, its symmetry reduces the calculation of $\lambda_2$ to a simpler eigenvalue problem.
This tree consists of a root and three identical binary branches, each with four depth layers containing \(1,2,4,\) and \(8\) vertices, respectively.
Since $L$ is real symmetric, Cauchy's interlacing theorem identifies $\lambda_2$ with the smallest eigenvalue of a branch block obtained by deleting the root row and column.
Symmetry within each branch further reduces this eigenvalue problem to four depth layers.
Then, we can show $\lambda_2=\lambda_3=\lambda_{\min}(B_4)$, where
\begin{equation*}
 B_4=\begin{pmatrix}
 3&-\sqrt2&0&0\\
 -\sqrt2&3&-\sqrt2&0\\
 0&-\sqrt2&3&-\sqrt2\\
 0&0&-\sqrt2&1
 \end{pmatrix}.
 \label{eq:uniform-graph-tree-block}
\end{equation*}
This eigenvalue is the smallest root of
 $\lambda^4-10\lambda^3+30\lambda^2-26\lambda+1=0$, yielding $\lambda_2\approx 0.0403$.

When such a symmetry reduction is unavailable, a graph cut provides a simple upper bound on $\lambda_2$, hence a sufficient condition for instability.
For an unweighted graph and a nonempty vertex subset $S\subset\mathcal{S}$, let $\partial S$ denote the set of edges joining $S$ to $S^{\mathrm c}\coloneqq \mathcal{S}\setminus S$.
A standard cut bound~\cite[Lemma 13.7.1]{Godsil2001} gives
\begin{equation*}
 \lambda_2\le\Lambda_S\coloneqq\frac{n\,|\partial S|}{|S|(n-|S|)}.
 \label{eq:uniform-graph-cut-bound}
\end{equation*}
Here $\Lambda_S$ measures the connectivity across the cut relative to the sizes of its two parts, with fewer crossing edges and a more balanced partition giving a tighter upper bound.
For the tree used in Sec.~\ref{sec:numexp}, this bound provides a conservative estimate of the instability threshold.
A minimum cut removes one edge incident to the root, separating a 15-vertex branch from the remaining 31 vertices.
This gives $\lambda_2\le\Lambda_S=46/(15\cdot31)\simeq0.098925$.
By Proposition~\ref{prop:uniform-graph-critical-A}, this bound yields a sufficient condition for instability: $A>1.091862$.

%% file: AppendixB-Deriv-Limit.tex
\section{Derivation of cue dynamics, continuous limits and timescales of the coupled dynamics}
\subsection{Derivation of cue modulation dynamics}
\label{sec:algorithm-derivation}

We derive the endogenous cue modulation dynamics~\eqref{eq:dist-G-learnSMFG-diff} as an approximate gradient descent on an instantaneous cost function.
Suppose agents update $V_t$ to learn optimal state value $V^*_{\mu_t}$ for the current population density $\mu_t$.
At each time step, the reward $r_g(\cdot\,;\mu_t)$ then prescribes a spatial reward landscape that determines the current optimal preferability $V^*_{\mu_t}$, as in our previous work on target-locating problem~\cite{Kato2025PRXLife}. 
Thus, we follow the same construction while allowing $\mu_t$ to evolve between updates.

For the linear mapping $V_t=\mathcal{F}[Z_t]=Z_t/\beta$ used in the main text, the cue field corresponding to the optimal value is $Z^*_{\mu_t}\coloneqq\beta V^*_{\mu_t}$.
We introduce an instantaneous cost function $\mathcal{C}_\mathcal{F}[Z_t;\mu_t]$ that measures the deviation of $Z_t$ from $Z^*_{\mu_t}$ and incorporates endogenous cue diffusion:
\begin{equation}
    \label{eq:distZ-diff-cost}
        \mathcal{C}_\mathcal{F}[Z_t;\mu_t]\coloneqq \frac{1}{2}\sum_{s\in \mathcal{S}}\mu_t(s)(Z^*_{\mu_t}(s)-Z_t(s))^2+\frac{D}{2\alpha}\sum_{s,s'\in\mathcal{S}} L_{ss'}Z_t(s)Z_t(s'),
\end{equation}
where $D \in \mathbb{R}_+$ represents the diffusion coefficient, 
$\alpha \in (0,1)$ is the rate of local cue modulation, and $L$ denotes the graph Laplacian of the environment $\mathcal{G}$.
Eq.~\eqref{eq:distZ-diff-cost} represents that diffusion is relatively stronger at vertices with fewer agents.

To describe cells that modulate the endogenous cue while moving, we focus on simultaneous updates of the value estimate and population density, as introduced in SM Sec.~\ref{sec:entropy-SMFG}.
At each time step, the population density evolves according to Eq.~\eqref{eq:mean-field-evolution}, while agents locally degrade or produce the endogenous cue to make the cue field $Z_t$ closer to the current optimal $Z^*_{\mu_t}$.
Reducing $\mathcal{C}_\mathcal{F}[Z_t;\mu_t]$ by gradient descent in \(Z_t\), with \(\mu_t\) treated as a parameter, would require the unknown optimal cue field $Z^*_{\mu_t}$, which cannot be determined from local observations alone.
Instead, following the semi-gradient descent in RL~\cite{BartoSutton2018}, we approximate the gradient by using the current estimate $Z_t$ and the immediate reward $r_g(\cdot\,;\mu_t)$,
leading to the following construction of the cue update~\eqref{eq:dist-G-learnSMFG-diff}:
\begin{subequations}    \label{eq:Gdiff-deriv}
    \begin{alignat}{3}
        Z_{t+1}(s)-Z_{t}(s)&=&&-\alpha \nabla_{Z}\mathcal{C}_\mathcal{F}[Z_t;\mu_t](s)\\
        &=&&-\alpha \mu_t(s)(Z_t(s)-Z^*_{\mu_t}(s))-D\sum_{s'\in \mathcal{S}}L_{ss'}Z_t(s')\\
        &\approx &&-\alpha\mu_t(s)\left(Z_t(s)-\beta r_g(s;\mu_t)-\ln\sum_{s'\in\mathcal{S}} p(s'|s)\exp(\gamma Z_t(s'))\right)\notag\\
        &&&-D\sum_{s'\in \mathcal{S}}L_{ss'}Z_t(s')\label{eq:Gdiff-step2}
    \end{alignat}
\end{subequations}
for $s\in\mathcal{S}$.
In Eq.~\eqref{eq:Gdiff-step2}, we approximate the unknown target $Z^*_{\mu_t}=\beta\mathcal{B}^*_{\mu_t}[V^*_{\mu_t}]$ with $\beta\mathcal{B}^*_{\mu_t}[Z_t/\beta]$, computed from the locally observed reward $r_g(s;\mu_t)$ and the neighboring cue field $Z_t$.
Here, $\mathcal{B}^*_{\mu_t}$ is the Bellman optimality operator~\eqref{eq:ermdp-foundation-optimality-operator} with reward $r_g(\cdot;\mu_t)$.

The derivation above can be extended to obtain other forms of cue modulation dynamics.
First, the linear relation between cue concentration and value estimate can be replaced by a continuous bijective mapping $V_t=\mathcal{F}[Z_t]$, with a corresponding target $Z^*_{\mu_t}=\mathcal{F}^{-1}[V^*_{\mu_t}]$.
Changing this mapping or introducing an additional bijective transformation in the update yields other forms of cue modulation dynamics, as described in Appendix~B of our previous work~\cite{Kato2025PRXLife}.
Second, the cue update could depend on past local population density through $r_g(\cdot;\mu_{t-k})$ to account for delays in the chemical signals that convey density information to cells.
Since past and current densities coincide at stationarity, the original stationary states may remain attainable.

\subsection{Derivation of the continuous limits}
\label{sec:continuous-limit}
We derive the continuous limit of the coupled dynamics~\eqref{eq:dist-G-learning-policy}, \eqref{eq:dist-G-learnSMFG-diff} on a one-dimensional lattice, with local reward functions $r_g(s;\mu)=r(s;\mu(s))$ and a density-dependent intrinsic movement policy.
The intrinsic transition probabilities depend on the population density at the destination:
\begin{subequations}    
    \label{eq:lazy-rw}
    \begin{align}
    p(x\pm h|x,\mu)&=\frac{1}{2}\sigma(\mu(x\pm h)),\\
    p(x|x,\mu)&=1-\frac{1}{2}\left[\sigma(\mu(x+h))+\sigma(\mu(x-h))\right],
    \end{align}
\end{subequations}
where $\sigma$ is a smooth function satisfying $0\leq\sigma\leq1$.
Typical choices for $\sigma$ are $\sigma(\mu(s))=[1-\mu(s)/{\mu_\textrm{max}}]_+$ and $\sigma(\mu(s))=\exp(-\mu(s)/\mu_\textrm{max})$~\cite{Painter2002CanApplMathQuart,Bubba2019JTheorBiol}.
Both choices decrease with $\mu(s)$, suppressing movement into more crowded sites and representing a local volume-filling effect.
The continuous equations~\eqref{eq:lin-continuous} in the main text are recovered when the intrinsic movement policy is density-independent.

To take the continuous limit, we consider a one-dimensional lattice $S^h\coloneqq \{0,\pm h,\pm 2h,\ldots\}$ with uniform spacing $h>0$ and time step $\Delta t^h=h^2$.
On this lattice, we denote the population fractions and endogenous cue concentrations under the discrete updates~\eqref{eq:mean-field-evolution} and~\eqref{eq:dist-G-learnSMFG-diff} by $\mu_t^h(x)$ and $Z_t^h(x)$, respectively.
We represent the corresponding population density per unit length $\mu_t^h(x)/h$ by a smooth field $\mu:\mathbb{R}_{\ge0}\times\mathbb{R}\to\mathbb{R}$ satisfying $\mu(t,x)=\mu_t^h(x)/h$ for $t\ge0$ and $x\in S^h$, so that $\int_{x-h/2}^{x+h/2}\mu(t,y)\dd{y}=\mu_t^h(x)+O(h^3)$.
We represent the endogenous cue concentration by a smooth field $Z:\mathbb{R}_{\ge0}\times\mathbb{R}\to\mathbb{R}$ satisfying $Z(t,x)=Z_t^h(x)$.
We scale the function $\sigma$ in Eq.~\eqref{eq:lazy-rw} and the corresponding intrinsic movement probabilities as
\begin{align*}
    \sigma^h(\mu^h(x\pm h))&=\frac{\Delta t^h}{h^2}\bar{\sigma}\left(\frac{\mu^h(x\pm h)}{h}\right),\\
    p^h_{\Delta t^h}(x\pm h|x,\mu^h)&=\frac{1}{2}\sigma^h(\mu^h(x\pm h)).
\end{align*}
For the calculations below, we write $\sigma^{h,\pm}(x,\mu^h)\coloneqq\sigma^h(\mu^h(x\pm h))/\Delta t^h=\bar{\sigma}(\mu^h(x\pm h)/h)/h^2$.
We also scale the rates of local cue modulation and cue diffusion as $\alpha^h_{\Delta t^h}=\bar{\alpha}\Delta t^h/h$ and $D^h_{\Delta t^h}=\bar{D}\Delta t^h/h^2$, respectively.
The local reward is scaled as $r^h_g(s;\mu^h(s))=\bar{r}_g(s;\mu^h(s)/h)$ with $\bar{r}_g:\mathbb{R}\times\mathbb{R}_{\ge0}\to\mathbb{R}$.

We first derive the continuous limit of the population update by expanding the cue-dependent transition probabilities.
Under the policy~\eqref{eq:dist-G-learning-policy}, the population evolves over one time step $\Delta t^h$ as
\begin{equation}
    \label{eq:master-discrete-discrete}
    \mu^h_{t+\Delta t^h}(x)=\sum_{y\in\{x,x\pm h\}}\mu^h_{t}(y)\pi^h_{\Delta t^h}[Z^h_t,\mu^h_t](x|y),
\end{equation}
where 
\begin{subequations}
    \begin{align*}
        \pi^h_{\Delta t^h}[Z^h,\mu^h](x|x)&\coloneqq(1-(\sigma^{h,+}(x,\mu^h)+\sigma^{h,-}(x,\mu^h))\Delta t^h/2)\exp(\gamma Z^h(x))/\mathcal{Z}^h_{\Delta t^h}[Z^h](x,\mu^h)\\
        \pi^h_{\Delta t^h}[Z^h,\mu^h](x\pm h|x)&\coloneqq\frac{\sigma^{h,\pm}(x,\mu^h)\Delta t^h}{2}\exp(\gamma Z^h(x\pm h))/\mathcal{Z}^h_{\Delta t^h}[Z^h](x,\mu^h)\\
        \mathcal{Z}^h_{\Delta t^h}[Z^h](x,\mu^h)&\coloneqq (1-(\sigma^{h,+}(x,\mu^h)+\sigma^{h,-}(x,\mu^h))\Delta t^h/2)\exp(\gamma Z^h(x))\notag\\
        &+\frac{\Delta t^h}{2}\left(\sigma^{h,+}(x,\mu^h)\exp(\gamma Z^h(x+h))+\sigma^{h,-}(x,\mu^h)\exp(\gamma Z^h(x- h))\right).
    \end{align*}
\end{subequations}
For the spatial Taylor expansions below, we suppress the time argument of the continuous fields, writing $\mu(x)=\mu(t,x)$ and $Z(x)=Z(t,x)$, and use the shorthand
    \(\bar{\sigma}(x)\equiv\bar{\sigma}(\mu(t,x)),
    \bar{\sigma}_x(x)=\bar{\sigma}'(\mu(t,x))\mu_x(t,x).\)
We first expand the terms entering the population update:
\begin{align*}
    \frac{\mu_t^h(x\pm h)}{h}&=\mu(x)\pm h\mu_x(x)+\frac{h^2}{2}\mu_{xx}(x)+O(h^3),\\
    \sigma^{h,\pm}(x,\mu_t^h)&=\frac{1}{h^2}\bar{\sigma}\left(\frac{\mu_t^h(x\pm h)}{h}\right)\\
    &=\frac{1}{h^2}\left(\bar{\sigma}(x)\pm h\bar{\sigma}_x(x)+\frac{h^2}{2}\bar{\sigma}_{xx}(x)+O(h^3)\right),\\
    
    Z_t^h(x\pm h)&=Z(x)\pm hZ_x(x)+\frac{h^2}{2}Z_{xx}(x)+O(h^3),\\
    \frac{\exp(\gamma Z_t^h(x\pm h))}{\exp(\gamma Z_t^h(x))} &=1\pm h\gamma\partial_xZ(x)+\frac{h^2}{2}(\gamma\partial_x^2Z(x)+(\gamma\partial_xZ(x))^2)+O(h^3),\\
    \sigma^{h,\pm}(x,\mu_t^h)\left(\frac{\exp(\gamma Z_t^h(x\pm h))}{\exp(\gamma Z_t^h(x))}-1\right)&=\frac{1}{h^2}\left(\bar{\sigma}(x)\pm h\bar{\sigma}_x(x)+\frac{h^2}{2}\bar{\sigma}_{xx}(x)+O(h^3)\right)\\
    &\left(\pm h\gamma\partial_xZ(x)+\frac{h^2}{2}(\gamma\partial_x^2Z(x)+(\gamma\partial_xZ(x))^2)+O(h^3)\right)\\
    &=\pm\frac{\bar{\sigma}(x)\gamma\partial_xZ(x)}{h}\\
    &\quad+\bar{\sigma}_x(x)\gamma\partial_xZ(x)+\frac{1}{2}\bar{\sigma}(x)(\gamma \partial_x^2 Z(x)+(\gamma\partial_xZ(x))^2)+O(h),
\end{align*}
These expansions give us the normalization factor $\mathcal{Z}^h_{\Delta t^h}[Z_t^h](x,\mu_t^h)$:
\begin{align}
    \mathcal{Z}^h_{\Delta t^h}[Z_t^h](x,\mu_t^h)&=\exp(\gamma Z_t^h(x))\Biggl[1+\frac{\Delta t^h}{2}\Biggl\{\sigma^{h,+}(x,\mu_t^h)\left(\frac{\exp(\gamma Z_t^h(x+h))}{\exp(\gamma Z_t^h(x))}-1\right)\notag\\
    &\qquad+\sigma^{h,-}(x,\mu_t^h)\left(\frac{\exp(\gamma Z_t^h(x- h))}{\exp(\gamma Z_t^h(x))}-1\right)\Biggr\}\Biggr]\notag\\
    &=\exp(\gamma Z_t^h(x))\left[1+\frac{\Delta t^h}{2}\left(2\bar{\sigma}_x(x)\gamma\partial_xZ(x)+\bar{\sigma}(x)(\gamma \partial_x^2 Z(x)+(\gamma\partial_xZ(x))^2)\right)+O(\Delta t^h h)\right].
    \label{eq:Z-normalization-expansion}
\end{align}
Using these expansions, we obtain the transition probabilities:
\begin{align*}
    \pi^h_{\Delta t^h}[Z_t^h,\mu_t^h](x\pm h|x)&=\frac{\sigma^{h,\pm}(x,\mu_t^h)\Delta t^h}{2}\exp(\gamma Z_t^h(x\pm h))/\mathcal{Z}^h_{\Delta t^h}[Z_t^h](x,\mu_t^h)\\
    &=\frac{\sigma^{h,\pm}(x,\mu_t^h)\Delta t^h}{2}\frac{\exp(\gamma Z_t^h(x))}{\mathcal{Z}^h_{\Delta t^h}[Z_t^h](x,\mu_t^h)}\frac{\exp(\gamma Z_t^h(x\pm h))}{\exp(\gamma Z_t^h(x))}\\
    &=\frac{\Delta t^h}{2h^2}\left(\bar{\sigma}(x)\pm h\bar{\sigma}_x(x)+O(h^2)\right)(1\pm h\gamma Z_x(x)+O(h^2))\\
    &\left[1-\frac{\Delta t^h}{2}\left(2\bar{\sigma}_x(x)\gamma\partial_xZ(x)+\bar{\sigma}(x)(\gamma \partial_x^2 Z(x)+(\gamma\partial_xZ(x))^2)\right)+O(\Delta t^h h)\right]\\
    &=\frac{\Delta t^h}{2h^2}\left(\bar{\sigma}(x)\pm h(\bar{\sigma}_x(x)+\gamma\bar{\sigma}(x)Z_x(x))+O(h^2)\right)\\
    &\left[1-\frac{\Delta t^h}{2}\left(2\bar{\sigma}_x(x)\gamma\partial_xZ(x)+\bar{\sigma}(x)(\gamma \partial_x^2 Z(x)+(\gamma\partial_xZ(x))^2)\right)+O(\Delta t^h h)\right],\\
    \pi^h_{\Delta t^h}[Z_t^h,\mu_t^h](x|x)&=\left(1-\frac{\Delta t^h}{2}(\sigma^{h,+}(x,\mu_t^h)+\sigma^{h,-}(x,\mu_t^h))\right)\frac{\exp(\gamma Z_t^h(x))}{\mathcal{Z}^h_{\Delta t^h}[Z_t^h](x,\mu_t^h)}\\
    &=\left(1-\frac{\Delta t^h}{2h^2}(2\bar{\sigma}(x)+h^2\bar{\sigma}_{xx}(x)+O(h^3))\right)\\
    &\left[1-\frac{\Delta t^h}{2}\left(2\bar{\sigma}_x(x)\gamma\partial_xZ(x)+\bar{\sigma}(x)(\gamma \partial_x^2 Z(x)+(\gamma\partial_xZ(x))^2)\right)+O(\Delta t^h h)\right].
\end{align*}
With $\Delta t^h=h^2$, these expansions simplify to
\begin{align*}
    \pi^h_{\Delta t^h}[Z_t^h,\mu_t^h](x\pm h|x)&=\frac{\bar{\sigma}(x)}{2}\pm\frac{h}{2}(\bar{\sigma}_x(x)+\gamma \bar{\sigma}(x)Z_x(x))+O(h^2),\\
    \pi^h_{\Delta t^h}[Z_t^h,\mu_t^h](x|x)&=1-\bar{\sigma}(x)+O(h^2).
\end{align*}
Using $q(x)=\bar{\sigma}_x(x)+\gamma\bar{\sigma}(x)Z_x(x)$ for brevity, we write the population update~\eqref{eq:master-discrete-discrete} in terms of incoming and outgoing transitions and Taylor expand the incoming terms about $x$:
\begin{align*}
    \frac{\mu^h_{t+\Delta t^h}(x)-\mu_t^h(x)}{h}
    &=\sum_{\eta=\pm1}\Bigl[\mu(x-\eta h)\pi^h_{\Delta t^h}[Z_t^h,\mu_t^h](x|x-\eta h)
    -\mu(x)\pi^h_{\Delta t^h}[Z_t^h,\mu_t^h](x+\eta h|x)\Bigr]\\
    &=-h\partial_x\left[\mu(x)\sum_{\eta=\pm1}\eta\pi^h_{\Delta t^h}[Z_t^h,\mu_t^h](x+\eta h|x)\right]\\
    &\quad+\frac{h^2}{2}\partial_x^2\left[\mu(x)\sum_{\eta=\pm1}\pi^h_{\Delta t^h}[Z_t^h,\mu_t^h](x+\eta h|x)\right]+O(h^3)\\
    &=h^2\left(\frac{1}{2}(\mu(x)\bar{\sigma}(x))_{xx}-(\mu(x)q(x))_x\right)+O(h^3).
\end{align*}
Restoring the time argument and dividing by $\Delta t^h=h^2$ gives
\begin{align*}
    \frac{\mu(t+\Delta t^h,x)-\mu(t,x)}{\Delta t^h}&=\frac{1}{2}(\mu(t,x)\bar{\sigma}(\mu(t,x)))_{xx}-(\mu(t,x)(\bar{\sigma}'(\mu(t,x))\mu_x(t,x)+\gamma\bar{\sigma}(\mu(t,x))Z_x(t,x)))_x+O(h)\\
    &=\frac{1}{2}(\bar{\sigma}(\mu(t,x))\mu_{xx}(t,x)-\mu(t,x)\partial_x^2[\bar{\sigma}(\mu(t,x))])-\gamma \partial_x(\mu(t,x)\bar{\sigma}(\mu(t,x))Z_x(t,x))+O(h)\\
    &=\frac{1}{2}\partial_x\left[(\bar{\sigma}(\mu(t,x))-\mu(t,x)\bar{\sigma}'(\mu(t,x)))\partial_x\mu(t,x)\right]-\gamma \partial_x(\mu(t,x)\bar{\sigma}(\mu(t,x))Z_x(t,x))+O(h).
\end{align*}
Taking $h\to0$, we formally obtain the following equation for the population density: 
\begin{equation}
    \partial_t\mu(t,x)=\frac{1}{2}\partial_x\left[\left(\bar{\sigma}(\mu(t,x))-\mu(t,x)\bar{\sigma}'(\mu(t,x))\right)\partial_x\mu(t,x)\right]-\gamma\partial_x\left[\mu(t,x)\bar{\sigma}(\mu(t,x))\partial_x Z(t,x)\right].
\end{equation}
For constant $\bar{\sigma}$, this reduces to Eq.~\eqref{lin-continuous-popu}.
This continuous limit shows that the exponentially biased volume-filling movement policy~\eqref{eq:dist-G-learning-policy} yields an advection–diffusion equation with advection along the cue gradient~\cite{Painter2002CanApplMathQuart}.

We next derive the cue equation under the same space-time scaling.
Over one time step $\Delta t^h$, the cue update~\eqref{eq:dist-G-learnSMFG-diff} at $x\in S^h$ is as follows:
\begin{align}
    Z^h_{t+\Delta t^h}(x)-Z^h_t(x)=&-\alpha^h_{\Delta t^h}\mu^h_t(x)\left(Z^h_t(x)-\beta r^h_g(x;\mu^h_t(x))-\left(\ln\sum_{y\in\{x,x\pm h\}}p^h_{\Delta t^h}(y|x,\mu_t^h)\exp(\gamma Z^h_t(y))\right)\right)\notag\\
    &+D_{\Delta t^h}^h\left(Z^h_t(x+h)-Z^h_t(x)+Z^h_t(x-h)-Z^h_t(x)\right).    \label{eq:dist-G-learning-discrete-discrete}
\end{align}
We use the expansion of the normalization factor in Eq.~\eqref{eq:Z-normalization-expansion} to obtain
\begin{equation*}
\begin{aligned}
    &\sum_{y\in\{x,x\pm h\}}p_{\Delta t^h}^h(y|x,\mu_t^h)\exp(\gamma Z_t^h(y))\\
    &\quad=\exp(\gamma Z_t^h(x))\left[1+\frac{\Delta t^h}{2}(\bar{\sigma}(x)(\gamma Z_{xx}(x)+(\gamma Z_x(x))^2)+2\gamma \bar{\sigma}_x(x)Z_x(x))+O(\Delta t^h h)\right].
\end{aligned}
\end{equation*}
The logarithm of this term is therefore \(\gamma Z(t,x)+O(h^2)\). 
Expanding Eq.~\eqref{eq:dist-G-learning-discrete-discrete} in $h$ and inserting the scalings of $\alpha^h_{\Delta t^h}$, $D^h_{\Delta t^h}$ and $r^h_g$ gives
\begin{subequations}
    \begin{align*}
        \frac{1}{\Delta t^h}\left(Z^h_{t+\Delta t^h}(x)-Z^h_t(x)\right)
        &=-\frac{\alpha^h_{\Delta t^h}}{\Delta t^h}h\mu(t,x)\left(Z(t,x)-\beta r^h_g(x;\mu^h_t(x))-\gamma Z(t,x)\right)+O(\frac{\alpha^h_{\Delta t^h}}{\Delta t^h}h^3)\notag\\
        &+\bar{D}\left(\partial_x^2Z(t,x)+O(h^2)\right)\\
        &=-\bar{\alpha}\left(\mu(t,x)\left(Z(t,x)-\beta \bar{r}_g(x;\mu(t,x))-\gamma Z(t,x)\right)\right)+\bar{D}\partial_x^2Z(t,x)+O(h^2).
    \end{align*}
\end{subequations}
Taking $h\to 0$, we formally obtain the following cue equation: 
\begin{equation}
\label{eq:continuous-dist-G-leaning-diff}
    \pt_t Z(t,x)=-\bar{\alpha}(1-\gamma) \mu(t,x) Z(t,x)+\bar{\alpha}\beta\mu(t,x) \bar{r}_g(x;\mu(t,x))+\bar{D}\nabla^2 Z(t,x).
\end{equation}
The scaling \(\alpha^h_{\Delta t^h}=\bar{\alpha}\Delta t^h/h\) compensates for the population fraction per site \(\mu_t^h(x)\simeq h\mu(t,x)\), allowing local cue modulation to survive alongside cue diffusion with \(D^h_{\Delta t^h}=\bar D\Delta t^h/h^2\) in the continuous limit.
This yields the cue dynamics of Keller-Segel type models.
However, with \(\gamma\) held fixed, the same limit reduces $\ln\sum p(y|x)\exp(\gamma Z(y))$ to \(\gamma Z\), eliminating its spatial coupling between states.
The local modulation consequently relaxes \(Z\) toward the rescaled instantaneous reward \(\beta\bar r_g/(1-\gamma)\), losing the propagation of discounted future rewards encoded by the Bellman equation~\cite{Todorov2009PNAS}.

\subsection{Relative timescales of population movement and cue modulation}
\label{sec:timescales-revised-draft}
The numerical experiments compare population movement and cue modulation by varying $\alpha$ relative to $\varepsilon$ while keeping $D/\alpha$ fixed.
To identify how these parameters govern the updates, we introduce a small time step $\Delta\tau$ while retaining the fixed graph $\mathcal{G}$.
We write $\varepsilon=\bar{\varepsilon}\Delta\tau$, $\alpha=\bar{\alpha}\Delta\tau$, and $D=\bar D\Delta\tau$.
The intrinsic movement policy used in the numerical experiments is written as
\begin{equation*}
    \label{eq:timescales-draft-intrinsic}
    p_{\Delta\tau}(s'|s)=
    \begin{cases}
        \bar{\varepsilon}\Delta\tau, & s'\in\mathcal{N}(s),\\
        1-\bar{\varepsilon}\Delta\tau|\mathcal{N}(s)|, & s'=s,\\
        0, & \text{otherwise}.
    \end{cases}
\end{equation*}

To leading order in \(\Delta\tau\), the population update depends linearly on \(\bar{\varepsilon}\), while the cue update depends linearly on \(\bar{\alpha}\) and \(\bar D\), as shown below.
Writing $A[Z_t](s';s)\coloneqq\exp(\gamma[Z_t(s')-Z_t(s)])$, we expand the normalizing denominator to first order in $\Delta\tau$, as in Eq.~\eqref{eq:Z-normalization-expansion}, to obtain
\begin{subequations}
    \label{eq:timescales-draft-policy}
    \begin{align*}
        \pi_{\Delta\tau}[Z_t](s'|s)&=\bar{\varepsilon}\Delta\tau A[Z_t](s';s)+O((\Delta\tau)^2),\\
        \pi_{\Delta\tau}[Z_t](s|s)&=1-\bar{\varepsilon}\Delta\tau\sum_{s'\in\mathcal{N}(s)}A[Z_t](s';s)+O((\Delta\tau)^2).
    \end{align*}
\end{subequations}
Substituting these probabilities into the population update~\eqref{eq:mean-field-evolution} gives
\begin{equation*}
    \label{eq:timescales-draft-population}
    \frac{\mu_{t+\Delta\tau}(s)-\mu_t(s)}{\Delta\tau}
    =\bar{\varepsilon}\sum_{s'\in\mathcal{N}(s)}\left[\mu_t(s')A[Z_t](s;s')-\mu_t(s)A[Z_t](s';s)\right]+O(\Delta\tau).
\end{equation*}
The cue update~\eqref{eq:dist-G-learnSMFG-diff} with the same time step can be written as
\begin{subequations}
    \label{eq:timescales-draft-cue}
    \begin{align*}
        \frac{Z_{t+\Delta\tau}(s)-Z_t(s)}{\Delta\tau}
        &=\bar{\alpha}\mu_t(s)\Delta Z_{t,\Delta\tau}(s)-\bar D\sum_{s'\in\mathcal{S}}L_{ss'}Z_t(s'),\\
        \Delta Z_{t,\Delta\tau}(s)
        &\coloneqq-Z_t(s)+\beta r_g(s;\mu_t)+\ln\sum_{s'\in\mathcal{S}}p_{\Delta\tau}(s'|s)\exp(\gamma Z_t(s')).
    \end{align*}
\end{subequations}
When \(\varepsilon\) is sufficiently small for the first-order expansion to apply and \(D/\alpha\) is fixed, the ratio \(\alpha/\varepsilon=\bar{\alpha}/\bar{\varepsilon}\) therefore approximately characterizes the relative timescales of cue dynamics and population movement, 
motivating the parameter comparison in the numerical experiments.

%% file: AppendixC-SupplementForExperiments.tex
\section{Experimental details}
\subsection{Numerical search for a stationary social optimum}
\label{sec:stationary-mfc-numerics}

We numerically approximate a stationary social optimum by maximizing $\mathcal J[\pi;\mu^\pi]$, the objective of the stationary mean field control (SMFC) problem~\eqref{eq:mfg-foundation-social-objective}.
For the density-independent intrinsic movement policy $p$ used in the numerical experiments, we parameterize movement policies by assigning a scalar parameter $\theta(s)$ to each vertex $s$:
\begin{equation}
    \pi_\theta(s'|s) = \frac{p(s'|s)\exp(\theta(s'))}{\sum_{u\in\mathcal S}p(u|s)\exp(\theta(u))},
    \qquad \theta\in\mathbb R^{|\mathcal S|}.
    \label{eq:mfc-numerics-policy}
\end{equation}
We center $\theta$ to have zero mean, since a constant shift leaves the policy unchanged.
For every candidate $\theta$, we find the associated stationary distribution $\mu^{\pi_\theta}$ by solving the following linear system:
\begin{equation}
    \mu^{\pi_\theta}(s') = \sum_{s\in\mathcal S}\mu^{\pi_\theta}(s)\pi_\theta(s'|s),
    \qquad \sum_{s\in\mathcal S}\mu^{\pi_\theta}(s)=1.
    \label{eq:mfc-numerics-stationarity}
\end{equation}
Since the initial distribution is stationary, the objective is evaluated directly as
\begin{equation}
\begin{split}
    \mathcal J[\pi_\theta;\mu^{\pi_\theta}] = \frac{1}{1-\gamma}\sum_{s\in\mathcal S}\mu^{\pi_\theta}(s)\biggl[ r_g\bigl(s;\mu^{\pi_\theta}\bigr) {}-\frac{1}{\beta}\sum_{s':\,p(s'|s)>0}\pi_\theta(s'|s)\log\frac{\pi_\theta(s'|s)}{p(s'|s)}\biggr].
\end{split}
    \label{eq:mfc-numerics-objective}
\end{equation}
Although this optimization problem is nonconvex, making it difficult to find the social optimum, we can still find policies with high objective values using a black-box optimization method.

Since this optimization problem is nonconvex, we also need to choose initial estimates carefully to find policies with high objective values.
To construct initial estimates of the stationary distribution, we first choose candidate set sizes using the reward term in Eq.~\eqref{eq:mfc-numerics-objective}, and then select vertex sets based on the policy regularization term.
When the reward is large relative to the policy regularization, a useful initial estimate concentrates the population approximately uniformly on a subset, at a density near the value that maximizes the reward.
For the reward $r_g(s;\mu(s))=A(\mu(s)/K)^a/[1+(\mu(s)/K)^b]$ used in the numerical experiments, this maximum occurs at $x_{\mathrm{peak}}=K[a/(b-a)]^{1/b}$.
Since a uniform population on \(m\) vertices has density \(1/m\), we consider sets $S\subseteq\mathcal S$ of size $m\in\{\lfloor 1/x_{\mathrm{peak}}\rfloor,\lceil 1/x_{\mathrm{peak}}\rceil\}$.
For the parameter set used here, $1/x_{\mathrm{peak}}$ is close to an integer.

For a fixed $S$ and $s\in S$, the transition row supported on $S$ that minimizes its KL divergence from $p(\cdot|s)$ is $p(s'|s)/z_s(S)$ for $s'\in S$, where $z_s(S)=\sum_{u\in S}p(u|s)$.
Its minimum KL cost is $-\log z_s(S)$, so for uniform mass on $S$ the corresponding averaged row cost is $C(S)/m$, where
\begin{equation*}
    C(S)=-\sum_{s\in S}\log z_s(S).
    \label{eq:mfc-support-surrogate}
\end{equation*}
This is a surrogate for the policy regularization cost, since these transition rows need not preserve the uniform distribution on $S$.
We use \(C(S)\) to select candidate vertex sets while using the original objective~\eqref{eq:mfc-numerics-objective} to rank the resulting policy candidates.

To minimize $C(S)$ at fixed size $m$, we reduce the problem to a mixed-integer linear program (MILP) by introducing auxiliary variables.
Since the intrinsic transition probabilities from each vertex to its neighbors are equal, we write $d_s=|N(s)|$, $a_s=p(s|s)$, and $b_s=p(u|s)$ for $u\in N(s)$.
We introduce binary membership variables $x_s$, continuous variables $w_{su}$ for undirected edges $\{s,u\}\in E$, and binary indicators $y_{sd}$ for a selected vertex having $d$ selected neighbors.
Since $z_s(S)=a_s+b_s d$ at such a vertex, the logarithmic costs are fixed coefficients in the following MILP:
\begin{equation}
\begin{aligned}
    \min_{x,w,y}\quad &-\sum_{s\in\mathcal S}\sum_{d=0}^{d_s}\log(a_s+b_s d)y_{sd},\\
    \text{subject to}\quad &\sum_{s\in\mathcal S}x_s=m,\\
    &\sum_{d=0}^{d_s}y_{sd}=x_s,\qquad \sum_{d=0}^{d_s}d\,y_{sd}=\sum_{u\in N(s)}w_{su},\\
    &0\leq w_{su}\leq x_s,\qquad w_{su}\leq x_u,\qquad w_{su}\geq x_s+x_u-1,\\
    &x_s\in\{0,1\},\qquad y_{sd}\in\{0,1\}.
\end{aligned}
    \label{eq:mfc-support-milp}
\end{equation}
The edge constraints enforce $w_{su}=x_sx_u$, so $w_{su}=1$ only when both endpoints of edge $\{s,u\}$ belong to $S$.
We solve this MILP using HiGHS through JuMP~\cite{Dunning2017SIAMrev}, adding $\sum_{s\in S^{(j)}}x_s\leq m-1$ after each examined set $S^{(j)}$ to exclude that exact set from subsequent solves.
To avoid repeatedly exploring similar initial sets, we retain a candidate only if its collection of degree multisets, one for each connected component, differs from those of previously examined sets.
We continue this procedure until 10 initial sets have been obtained.

For each retained \(S\), we construct an initial policy whose stationary distribution approximates the uniform distribution on \(S\).
We first define a strictly positive initial density by distributing total mass \(1-\eta\) uniformly over \(S\) and mass \(\eta\) uniformly over its complement \(S^c\), where \(\eta\in\{0.001,0.005,0.01,0.02,0.05\}\):
$\rho_{S,\eta}(s)=(1-\eta)/|S|,\ s\in S$ and $\rho_{S,\eta}(s)=\eta/(|\mathcal S|-|S|),\ s\notin S$.
We then compute the centered parameters $\theta_{S,\eta}$ whose policy has invariant density $\rho_{S,\eta}$ using Eqs.~\eqref{eq:mfc-numerics-policy} and~\eqref{eq:mfc-numerics-stationarity}.
For each $S$, we compute $\mathcal J[\pi_{\theta_{S,\eta}};\mu^{\pi_{\theta_{S,\eta}}}]=\mathcal J[\pi_{\theta_{S,\eta}};\rho_{S,\eta}]$ for every $\eta$, and retain the value of $\eta$ that yields the highest objective value.

We maximize Eq.~\eqref{eq:mfc-numerics-objective} using a black-box optimization method, specifically the covariance matrix adaptation evolution strategy (CMA-ES), with a different initial estimate for each restart.
We rank these retained policies by their objective value $\mathcal J[\pi_{\theta_{S,\eta}};\mu^{\pi_{\theta_{S,\eta}}}]=\mathcal J[\pi_{\theta_{S,\eta}};\rho_{S,\eta}]$ and use them to initialize up to 10 CMA-ES restarts, with one policy per restart.
Each restart terminates at its generation limit, on convergence, or after 30 generations without an objective improvement greater than $10^{-8}$.
We denote the policy with the highest objective by $\pi^*_{\mathrm{best}}$  as in the main text.